\documentclass{emulateapj}
\usepackage{placeins}
\usepackage{graphicx}

\shorttitle{Transitional Disk Gaps and Planets} \shortauthors{Zhu et al.}

\newcommand\msun{\rm M_{\odot}}

\newcommand\msunyr{\rm M_{\odot}\,yr^{-1}}
\newcommand\be{\begin{equation}}
\newcommand\en{\end{equation}}

\newcommand\etal{{\rm et al}.\ }

\newcommand\mdot{\rm \dot{M}}

\begin{document}

\title{Transitional and Pre-Transitional disks: Gap Opening by Multiple Planets ?}

\author{Zhaohuan Zhu\altaffilmark{1},
Richard P. Nelson\altaffilmark{2},
Lee Hartmann\altaffilmark{1},
Catherine Espaillat\altaffilmark{3,4}, and Nuria Calvet\altaffilmark{1}}

\altaffiltext{1}{Dept. of Astronomy, University of Michigan, 500
Church St., Ann Arbor, MI 48109}
\altaffiltext{2}{Astronomy Unit, Queen Mary, University of London,
Mile End Road, London E1 4NS UK}
\altaffiltext{3}{Center for Astrophysics,
60 Garden St., Cambridge, MA 02138}
\altaffiltext{4}{NSF Astronomy \& Astrophysics Postdoctoral Fellow}

\email{zhuzh@umich.edu, lhartm@umich.edu, r.p.nelson@qmul.ac.uk, cespaillat@cfa.harvard.edu,
ncalvet@umich.edu}

\begin{abstract}
We use two-dimensional hydrodynamic simulations of viscous disks
to examine whether dynamically-interacting multiple giant planets
can explain the large gaps (spanning over one order of magnitude in radius) inferred for the transitional
and pre-transitional disks around T Tauri stars.
In the absence of inner disk dust depletion, we find that it requires three to four
giant planets to open up large enough gaps to be consistent with
inferences from spectral energy distributions,
because the gap width is limited by the tendency of the planets to be driven
together into 2:1 resonances.
With very strong tidal torques and/or rapid planetary accretion, fewer planets can also generate
a large cavity interior to the locally formed gap(s) by preventing outer disk
material from moving in. In these cases, however, the reduction of surface density produces
a corresponding reduction in the inner disk accretion rate onto the star;
this makes it difficult to explain the observed accretion
rates of the pre/transitional disks.
We find that even with four planets in disks, additional
substantial dust depletion is required to explain observed disk gaps/holes.
Substantial dust settling and growth, with consequent
significant reductions in optical depths, is inferred for
typical T Tauri disks in any case, and an earlier history of dust growth is consistent
with the hypothesis that pre/transitional disks are explained by
the presence of giant planets. We conclude that the depths and widths of gaps,
and disk accretion rates in pre/transitional disks cannot be reproduced by a
planet-induced gap opening scenario alone. Significant dust depletion is also required
within the gaps/holes. Order of magnitude estimates suggest the mass
of small dust particles ($\lesssim 1 \mu$m) relative to the gas must be depleted to
10$^{-5}$ -- 10$^{-2}$ of the interstellar medium value, implying a very efficient
mechanism of small dust removal or dust growth.

\end{abstract}


\section{Introduction}

The transitional and pre-transitional disks around young stars exhibit
strong dust emission at wavelengths $\gtrsim 10 \mu$m, while showing significantly
reduced fluxes relative to typical T Tauri disks at shorter wavelengths
(e.g., Calvet \etal 2002, 2005; D'Alessio \etal 2005;
Espaillat \etal 2007,  2008).  In the pre-transitional disks, there is evidence
for emission from warm, optically-thick dust near the star
(Espaillat \etal 2007, 2008, 2010), while in the transitional disks
the emission at $\lesssim 10 \mu$m appears to be due entirely to optically-thin
dust (Calvet \etal 2002, 2005; Espaillat \etal 2010).
The depletion
of near- to mid-infrared emission is generally interpreted as being due to
evacuation of the disk interior to scales $\sim 5$ to $\sim 50$~AU
(Marsh \& Mahoney 1992; Calvet \etal 2002, 2005; Rice \etal 2003;
Schneider \etal 2003; Espaillat \etal 2007, 2008, 2010; Hughes \etal 2009),
an interpretation confirmed in some cases via direct sub-mm imaging
(e.g., Pietu \etal 2006; Brown \etal 2007, 2009; Hughes \etal 2009; Andrews \etal 2009).

One proposed mechanism for clearing inner disks while leaving the outer
disk relatively undisturbed is the formation of
giant planets, which can open gaps in disks (e.g., Lin \& Papaloizou 1986;
Marsh \& Mahoney 1992; Nelson \etal 2000; Calvet \etal 2002; Rice \etal 2003),
and which are expected to form initially at relatively small disk radii due to
the shorter evolutionary timescales compared with the outermost
disk.  However, there are three observational challenges,
as discussed further in \S 2,
which theoretical explanations of these systems must confront.

First, the transitional and pre-transitional disk systems
exhibit average gas accretion rates close to T Tauri disk accretion rates
($\sim 10^{-8}$$\msunyr$; Hartmann et al. 1998) onto
their central stars (e.g. Calvet \etal 2002, 2005; Espaillat \etal 2007, 2008;
Najita et al. 2007).  Maintaining this accretion requires either a
significant mass reservoir interior to the disk-clearing planets,
or some way of allowing mass from the outer disk
to move past the gap-clearing planets.

Second, the cleared regions in these disks are {\em large}
(e.g. Espaillat \etal 2010).  In the case of
the transitional disks, the optically-thin region must extend from radii as
large as tens of AU all the way in to the central star.  Even the
pre-transitional disks, which have evidence for optically-thick dust emission
in the innermost regions, must have large disk gaps.
Furthermore, the spectral energy distribution (SED) modeling
suggests that the detected optically thick region may only extend
to radii $\sim 1$~AU, with an extremely dust free region
beyond, until reaching the outer optically-thick disk
(Espaillat \etal 2010; \S 2).

Third, the requirement that the gap/hole be optically thin implies that the
mass of dust in sizes of order a micron or less must be extremely small.
Thus, either the planet-induced gap is very deep and is effectively
cleared of gas {\it and} dust, or the dust abundance is reduced by many
orders of magnitude from abundances in the diffuse interstellar medium.

These three conditions must be fulfilled simultaneously.
Photoevaporation has been proposed as one gap/hole clearing
mechanism (Alexander \& Armitage 2007, 2009), but requires the mass loss
rate by the photoevaporation to be comparable to the disk accretion rate.
Initial estimates of photoevaporative mass loss suggested values of
$\sim$ 4$\times$10$^{-10}$$\msunyr$ (Clarke et al. 2001), which would
not be rapid enough to counteract accretion.
More recent estimates suggest higher values, perhaps as much as
$\sim$10$^{-8}$$\msunyr$ due to the inclusion of X-rays
(Gorti \& Hollenbach 2009; Owen et al. 2010); however, the highest
values are problematic, because then it becomes difficult to understand
why so many T Tauri disks last for several Myr with accretion rates smaller
than 10$^{-8}$$\msunyr$ (Hartmann et al. 1998).

Najita, Strom, \& Muzerolle (2007) suggested that giant planets might
explain the transitional disks by creating a gap while still maintaining accretion
onto the central star.  Najita \etal argued that there is some evidence of reduction
in the accretion rates in transitional disks relative to the so-called
``primordial'' disks, perhaps by an order of magnitude, and that this was consistent
with simulations of a single giant planet in the studies by
Lubow \& D'Angelo (2006) and Varniere \etal (2006).
However, as discussed in \S 2, the accretion rates of the pre/transitional disks
are not very low in absolute magnitude.  Moreover,
while Varniere \etal (2006) were able to produce a large cleared
inner region with a single planet, the study of
Crida \& Morbidelli (2007) calls this result into question (see also \S 3).

In this paper we consider whether disk gap formation by a multiple planet
system can satisfy the following requirements: 1) creation of a gap
extending over a large range in radius; 2) maintenance of the inner
disk accretion rate by flow through the planet-created gap; and
3) sufficient reduction in the disk surface density within the gap such
that extreme depletion of disk dust is not essential.
We examine the problem in the context of viscous disks, with improvements
over some previous simulations, including a more realistic temperature
distribution and an improved inner boundary condition.
We find that while multiple giant planets can indeed open large gaps, it
is difficult to explain the inferred properties of the pre/transitional
disks without invoking substantial depletion of small dust.  We offer
a speculative scenario in which combined dust depletion and planet formation
explains the observations.

\section{Inferred properties of pre/transitional disks}

We first present a summary of properties of some
of the known pre/transitional systems in Table 1 as derived from analysis of optical, infrared, and
millimeter SEDs, including
the pre-transitional disk systems LkCa 15, UX Tau A, Rox 44,
and of the transitional disk GM Aur.  The parameters are mostly
taken from Espaillat \etal (2010), where the methods used to derive
these estimates are discussed.

As indicated in Table 1, the estimated maximum extent of the inner
optically-thick region in the pre-transitional
disks is $\sim 0.2 - 0.4$~AU, while the inner edges of the
optically-thick outer disks lie at $\sim 40 - 70$~AU.
The location of the inner edge of the outer disk inferred from SED modeling is
consistent with that derived using millimeter interferometry (Pietu et
al. 2006; Andrews et al. 2009; S. Andrews, 2010, private communication).
The maximum extent of the inner disk is less certain and is based on SED
modeling of the near-infrared emission. In this wavelength region, the
flux in pre-transitional disks is dominated by the emission of the inner
wall. To limit the size of the optically thick disk behind the inner
wall, Espaillat et al. (2010) varied the maximum extent of this inner
disk and fitted it to the SED of the object. The value listed in Table 1
represents the largest R$_{inner disk}$ that does not produce too much
excess emission, assuming a typical disk flaring structure.
A caveat is that if the inner wall is ``puffed up''
relative to the disk behind it (e.g. Natta et al. 2001),
or if the dust behind the inner wall is more
settled than the wall, the inner disk would be in
shadow and not emit strongly.  However, we note that
the near-infrared interferometric observations by Pott \etal
(2010) are consistent with the dust emission being confined to a small
radius as inferred from the SED modeling.  The amount of {\em emitting}
optically thin sub-micron-sized dust is small, $\sim 10^{-11} M_{\odot}$.

Transitional disks do not have the inner optically thick disk observed
in the pre-transitional disks. The transitional disk of GM Aur has a 20
AU hole, based on modeling both the SED and submillimeter
interferometric visibilities (Hughes et al. 2009). About 10$^{-12}$
$M_{\odot}$ of sub-micron-sized optically thin dust is necessary within
the inner AU of the hole to reproduce the observed silicate emission in
the near-infrared.  Calvet \etal (2005) inferred that the inner
emitting, optically thin dust in GM Aur is confined to within
5 AU by modeling its SED.  While SED
modeling is nonunique, the interferometric observations of
Pott \etal (2010) indicated that this dust is confined within even
smaller radii, $\lesssim 0.15$~AU (see also Akeson \etal 2005).
Thus, as stated by Pott \etal (2010), the pre/transitional disks are
often characterized by distinct dust zones ``which are not smoothly
connected by a continuous distribution of optically thick material'',
which may be interpreted as the result of a highly cleared gap
with an inner region of optically thin or thick dust.

The final important point to be made is that, as shown in Table 1,
the gas accretion rates onto the central stars are substantial,
$\sim 0.3 - 1 \times 10^{-8} \msunyr$, not much lower than
rates typical of T Tauri stars (e.g., Hartmann \etal 1998).
These high gas accretion rates agree with significant
gas emission detected at the inner disk by \emph{Spitzer} (Najita \etal 2010). Thus, the depletion
of dust optical depth appears to be orders of magnitude larger than
any reduction in the gas accretion rate.  This poses strong constraints
on theories of the origin of pre/transitional disk structure, as we
now show.

\section{Methods: Disk--planet simulations}

To simulate multiple giant planet systems, we use FARGO (Masset 2000).
This is a two-dimensional hydrodynamic code which utilizes a fixed
grid in cylindrical polar coordinates ($R$, $\phi$).
FARGO uses finite differences to approximate derivatives, and the
evolution equations are divided into source and transport steps, similar to those of
ZEUS (Stone \etal 1992). However, an orbital advection scheme has
been incorporated which reduces the numerical diffusivity and significantly
increases the allowable timestep as limited by the
Courant-Friedrichs-Lewy (CFL) condition.
Thus FARGO enables us to study the interaction between the disk and embedded
planets  over a full viscous timescale for disks whose inner radii
are considerably smaller than the radial locations of the embedded planets,
which is an essential requirement for studying large gaps created by
multiple planet systems.

We assume a central star mass of $1 \msun$ and a fully viscous disk.
We further assume a radial temperature distribution
$ T = 221 (R/AU)^{-1/2}$~K, which is roughly consistent with typical T Tauri
disks in which irradiation from the central star dominates the disk temperature
distribution (e.g., D'Alessio \etal 2001). The disk is locally isothermal.
The adopted radial temperature distribution corresponds to an implicit ratio
of disk scale height to cylindrical radius $H/R = 0.029 (R/AU)^{0.25}$.
This differs from the $H/R =$~constant assumption used in many previous
simulations, which implies a temperature distribution
$T \propto R^{-1}$, which is inconsistent with observations. Consequently,
our assumed temperature distribution with a constant
viscosity parameter $\alpha$ ($\nu =\alpha c_{s}^{2}/\Omega$, where $\nu$ is the
kinematic viscosity, $c_s$ is the sound speed, $\Omega$ is the angular velocity)
leads to a steady-disk surface density distribution
$\Sigma \propto R^{-1}$ instead of the $\Sigma \propto R^{-1/2}$ which would
result from either assuming both $H/R $ and $\dot{M}$ are constant, or both $\nu$ and viscous torque (-2$\pi R\Sigma\nu R^{2}d\Omega/dR$) are constant.
This makes a significant
difference in the innermost disk surface densities, and thus the
implied inner disk optical depths in our models will be larger.

We set $\alpha = 0.01$ for the standard cases.
Given our assumed disk temperature distribution,
we set the initial disk surface density to be
$\Sigma$=178 (R/AU)$^{-1}$ (0.01/$\alpha$) g cm$^{-2}$
from $R \sim 1 - 200$~AU, which yields a steady
disk solution with an accretion rate $\dot{M} \sim 10^{-8} \msunyr$,
typical of T Tauri disks (Gullbring \etal 1998; Hartmann \etal 1998).
The mass within the radius $R_{out}$ is then
\begin{eqnarray}
M(R<R_{out})&=&\int_{R_{in}}^{R_{out}}2\pi R \Sigma(R) dR \nonumber\\
&=&1.26 \times 10^{-4} \frac{0.01}{\alpha} \left(\frac{R_{out}}{\rm AU}-\frac{R_{in}}{\rm AU}
\right)M_{\odot} \,.
\label{eq:diskmass}
\end{eqnarray}
With $R_{in}$=1 AU, $R_{out}$=200 AU, and $\alpha$=0.01, the total disk mass is thus
$0.025 M_{\odot}$, reasonably consistent with observational
estimates (e.g., Andrews \& Williams 2005).

We use logarithmic spacing for the radial grid to span the large
dynamic range needed.
The azimuthal resolution is 256 cells covering $2 \pi$ radians, and the
radial resolution is adjusted so that each grid cell is square
($R\Delta\theta = \Delta R$).

Each planet in the disk can interact gravitationally with the other planets and with the
disk. The interactions among multiple planets and the star are solved using
a 5th order Runge-Kutta integrator.
Each planet feels the acceleration from the disk given by
\begin{equation}
{\bf a}=-G \int_{S} \frac{\Sigma({\bf r}')({\bf r}_p - {\bf r})'  dS}{(|{\bf r}'-{\bf r}_{p}|^{2}+\epsilon^{2})^{3/2}}\,,
\end{equation}
where the integral is performed
over the whole disk area $S$. Here ${\bf r}_{p}$ denotes the planet's position,
and $\epsilon$ is the smoothing length set to be $0.6 H$, where $H$
is the disk scale height at the planet's orbit.

To accurately simulate close encounters between planets,
an additional timestep constraint is applied.  For each pair
of planets ($i, j$) we calculate the minimum timestep
\begin{equation}
dt = \frac{1}{400} min \left(\frac{ 2\pi}{\Omega_{ij}}\right)\,,
\label{orbit_ij}
\end{equation}
where $\Omega_{ij}= [ G(m_i + m_j)/|{\bf r}_i - {\bf r}_j|^3 ]^{1/2}$ is
an estimate of the orbital frequency that each pair
of planets (with masses $m_i$ and $m_j$) would have if they were
in orbit around each other,
with $|{\bf r}_i - {\bf r}_j|$ being the absolute value of the distance between them.
Throughout most of the simulations the time step size is determined
by the normal CFL condition, but when a close encounter occurs the
interaction is resolved by allowing Eq.~(\ref{orbit_ij}) to determine
the time step size.

Accretion onto the planet is simulated by depleting the disk surface density
within the planet's Hill radius, $R_H= (M_{P}/3 M_{*})^{1/3}r_{p}$,
where $M_P$ and $M_*$ are the planet and stellar masses, respectively.
This approach is required because of the necessity to evolve the disk-planet system for
longer than the viscous time scale measured at the disk outer edge, which forces
us to adopt a relatively low numerical resolution that prevents us from modeling
accurately the gas flow onto the planet within the planet Hill sphere.
Defining a dimensionless scaling parameter, $f$,
we deplete the disk within $0.45 R_H$ at a rate
\begin{equation}
\frac{d\Sigma}{dt}=-\frac{2 \pi}{fT_{P}}\Sigma\,,
\end{equation}
and between 0.45 and 0.75~$R_H$ at a reduced rate
\begin{equation}
\frac{d\Sigma}{dt}=-\frac{2 \pi}{3fT_{P}}\Sigma\,,
\end{equation}
where $T_P$ is the orbital period of the
planet\footnote{Note that our parametrization is different from the publicly
available version of FARGO.}.
We term $f$ the planet accretion timescale parameter.
During the intermediate stages of giant planet formation, after the formation of
the solid core, but prior to the onset of rapid gas accretion,
gas slowly settles onto the planet at a rate determined by the Kelvin-Helmholtz
time scale. Depending on the opacity of the planet envelope, this phase lasts for
up to a few Myr (Pollack et al. 1996; Papaloizou \& Nelson 2005;
Movshovitz et al. 2010). Once the planet reaches a value of $\sim 35$ -- 50 M$_{\oplus}$,
rapid gas accretion onto the planet is able to ensue, relatively unimpeded by
the thermodynamic evolution of the atmosphere, and the planet can grow quickly to
become a giant. during this final phase of growth, however, the planet contracts
rapidly, becoming much smaller than its Hill sphere, and further gas accretion
must occur via flow through a circumplanetary disk (Papaloizou \& Nelson 2005).
Simulations presented by Ayliffe \& Bate (2009) suggest that the circumplanetary disk
has a radius $R_{cp} \simeq R_H/3$ and aspect ratio $(H/R)_{cp} \simeq 0.5$.
Adopting these values, the viscous time scale for the circumplanetary
disk
\begin{equation}
\tau_{acc} \simeq 1/(5 \pi \alpha)T_{P}\, \equiv f/(2\pi),
\end{equation}
from which we see that $f = 2/(5 \alpha)$.
Varying the value of $\alpha$ between representative values of $10^{-2}$ and $10^{-3}$,
leads to variations of the accretion time of between $\sim 6$ and 60 planet orbits.
Since $f$ is poorly constrained, we vary it by two orders of magnitude in this work:
$1 \le f \le 100$.

Observations of exoplanets allow us to place some constraints on the value of $f$.
Once gap formation ensues, mass accretion into the planet Hill sphere is controlled by
the rate at which material is supplied to the planet by viscous evolution of
the protplanetary disc. A mass accretion rate of $10^{-8}$ M$_{\odot}$/yr implies
growth of the planet to masses $\ge 10$ M$_J$ over disc lifetimes of $\sim$ few Myr
if $f=1$. Given that exoplanets with masses $> 10$ M$_J$ are relatively rare,
this suggests that $f=1$ is a reasonable lower limit.
A value of $f=100$ implies mass accretion
onto the planet at the rate $\sim 10^{-9}$-$10^{-10}$ M$_{\odot}$/yr ($10^{-6}$-$10^{-7}$ M$_J$/yr).
Such low values of $f$ can barely explain the numerous exoplanets with masses in the range
1 - 5 M$_J$, suggesting that $f=100$ is a reasonable upper limit.

We initiate most of our simulations with relatively low planet masses
(details are given later in this section), and the fact that
planets grow slowly if $f=100$  means that during our simulations
rapid type I and type III migration are likely to dominate
the evolution of the planet orbits (Masset and Papaloizou 2003).
To avoid this, we initiate simulations with $f=100$ with the smaller value
$f=10$ until a gap opens and the planet reaches 1 $M_{J}$, at which point
$f$ is switched back to 100. The mass and momentum
of the depleted portion of the disk are added to the planet's mass and momentum,
such that the planet's migration will be (modestly) affected by its accretion from the disk.

As discussed by Crida, Morbidelli, \& Masset (2007),
a standard open inner boundary condition (Stone \etal 1992) in a fixed 2D grid can
produce an unphysically rapid depletion of material through
the inner boundary in the presence of the planets.
There are two reasons for this.  First, due to waves excited by the planet,
the gas in the disk can have periodic inward and
outward radial velocities larger than the net viscous velocity of accreting
material.
Thus, with the normal open boundary, material can
flow inward while there is no compensating outflow allowed.
Second, the orbit of the gas at the inner boundary is not circular
due to the gravitational potential of the planets;
again, as material cannot pass back out
through the inner boundary, rapid depletion of the inner
disk material is
enhanced.
As we are interested in the amount of gas depletion in the disk inward of the
planet-induced gap(s) over substantial evolutionary timescales,
it is important to avoid or minimize this unphysical mass depletion.

Crida \etal (2007) were able to ameliorate this problem by surrounding the
2D grid by extended 1D grids (see their Figure 5).  We follow Pierens \& Nelson
(2008), who found reasonable agreement with the Crida \etal results while
using a 2D grid only by limiting the inflow velocities at the inner boundary
to be no more than a factor $\beta$ larger
than the viscous radial velocity in a steady state,
\begin{equation}
v_{rs} = - \frac{3 \nu_{in}}{2R_{in}} \,,
\end{equation}
where $\nu_{in}$ and $R_{in}$ are the viscosity and radius at
the inner boundary. Pierens \& Nelson (2008) found satisfactory behavior
for $\beta = 5$ in their simulations, based on a comparison with 1D-2D
calculations.  As our parameters differ somewhat from
those of Pierens \& Nelson, we also made a comparison with the
1D-2D setup in FARGO described by Crida \etal (2007),
and found that $\beta = 3$ provided reasonable results.
We adopted the normal open boundary
condition as our outer boundary condition
\footnote{This is different from the publicly available version of FARGO.}.
The open outer boundary allows mass to leave
the computational domain, and in strongly perturbed disks containing
massive planets this can cause a noticeable reduction in the disk mass
over long time scales which in principle can reduce the gas accretion
rate through the disk. But because this effect requires a long time to develop,
its effect on our results is very modest.

We do not use the 1D-2D grid method in this paper because in
some cases our planet masses grow considerably larger than one Jupiter mass,
and the non-circular motion of the gas close to the 2D boundary is
significant enough so that an artificial gap is opened at the interface
between the 1D and 2D grids, which causes the code to crash.

Since our numerical inner boundary is far away from the real disk inner boundary,
we assign the ghost zone density to be
\begin{equation}
\Sigma_{g}=\Sigma_{1}\frac{R_{1}}{R_{g}}\,
\end{equation}
to simulate the power law section of the constant $\alpha$ disk similarity solution,
where subscript $g$ denotes the ghost zone and $1$ the first active zone.

In order to reduce problems related to relaxation from initial conditions,
and to mimic the effect of accreting gas onto a protoplanet at the beginning
of the rapid gas accretion phased discussed earlier in this section,
we performed the following steps. First, we allowed for the
planet mass to grow linearly from zero to its initial mass
during the first 5000 years. For simulations where the planet accretes
gas from the disc the initial mass is 0.1 M$_J$, and is 1 M$_J$ otherwise.
During this period the planet does not
feel the gravitational force from the disk and does not accrete gas.
This step is necessary; otherwise the sudden presence of a one Jupiter mass planet in a
Keplerian disk disturbs the disk so much that the excited acoustic waves
can be present in the disk for a long time \footnote{Waves can reflect back
from the boundary even with the open boundary condition.}, and
these waves gradually deplete the disk mass by creating flows through the outer
boundary.  Then we allow
the planet to start accreting at $10^4$ yr.
Finally, we allow the planet to feel the disk's gravitational force gradually,
starting at $10^4$ yr and with full force at $t = 2 \times 10^4$ yr. This gradual
ramping up of the gravitational effects is also important, since a sudden
turn-on of gravity can kick the planet outside the gap it just opened.
As well as being convenient from a computational point of view,
this approach also ensures that gap formation arises both because
of tidal torques and accretion onto the planet.

Problems with boundary conditions also led us to perform simulations
without the indirect terms in the gravitational potential
experienced by the disk and planets.  These terms arise when working in the
non--inertial reference frame based on the central star, as they
account for the acceleration of the central star by the disk and planets.
The resulting non-circular motions can cause problems at both inner and
outer boundaries of our circular grid similar to those discussed above.
Given the importance of these issues to our investigation of the
disk surface density distribution exterior to gaps, we therefore did
not include the indirect terms initially.
We subsequently recomputed the models P4AN and P4A10 (see following
section) to include the indirect
terms to ensure that the planetary stability properties and the gap structures
that we report in this paper remain effectively unchanged; we do indeed find
that this is the case.  (The result for the P4A10 case with indirect terms is shown in Fig. \ref{fig:AM2P4A10})
In performing these tests with the indirect terms
we found it necessary to adopt the wave-damping boundary condition described
in De Val-Borro \etal (2006) in order for the disk to remain well behaved
at the outside edge. Adopting this boundary condition has very little influence,
however, on the structure of the gap or the stability properties of the
planetary systems, so we believe that the results presented in this paper
provide an accurate description of the features most important for comparing with
observed pre/transitional disk structures.

\section{Results}

We performed simulations with 1, 2, 3 and 4 planets in
the disk to study planet gap opening and gas flowing through gaps.
The disk simulations evolve over 1 million years (Myr), which is
the characteristic viscous timescale at $R \sim 100$~AU,
and is also shorter than the time scale over
which most T Tauri disks are observed to clear.
The properties of the simulations are summarized in Table 2.

\subsection{Viscous disks with one planet}
For comparison and test purposes, we first calculated a case without planets (case PN).
The disk evolves as expected for a viscous disk with constant $\alpha$,
maintaining a surface density $\Sigma \propto R^{-1}$ and a nearly
constant mass accretion rate of 10$^{-8}$$\msunyr$.

\begin{figure}
\includegraphics[width=0.42\textwidth]{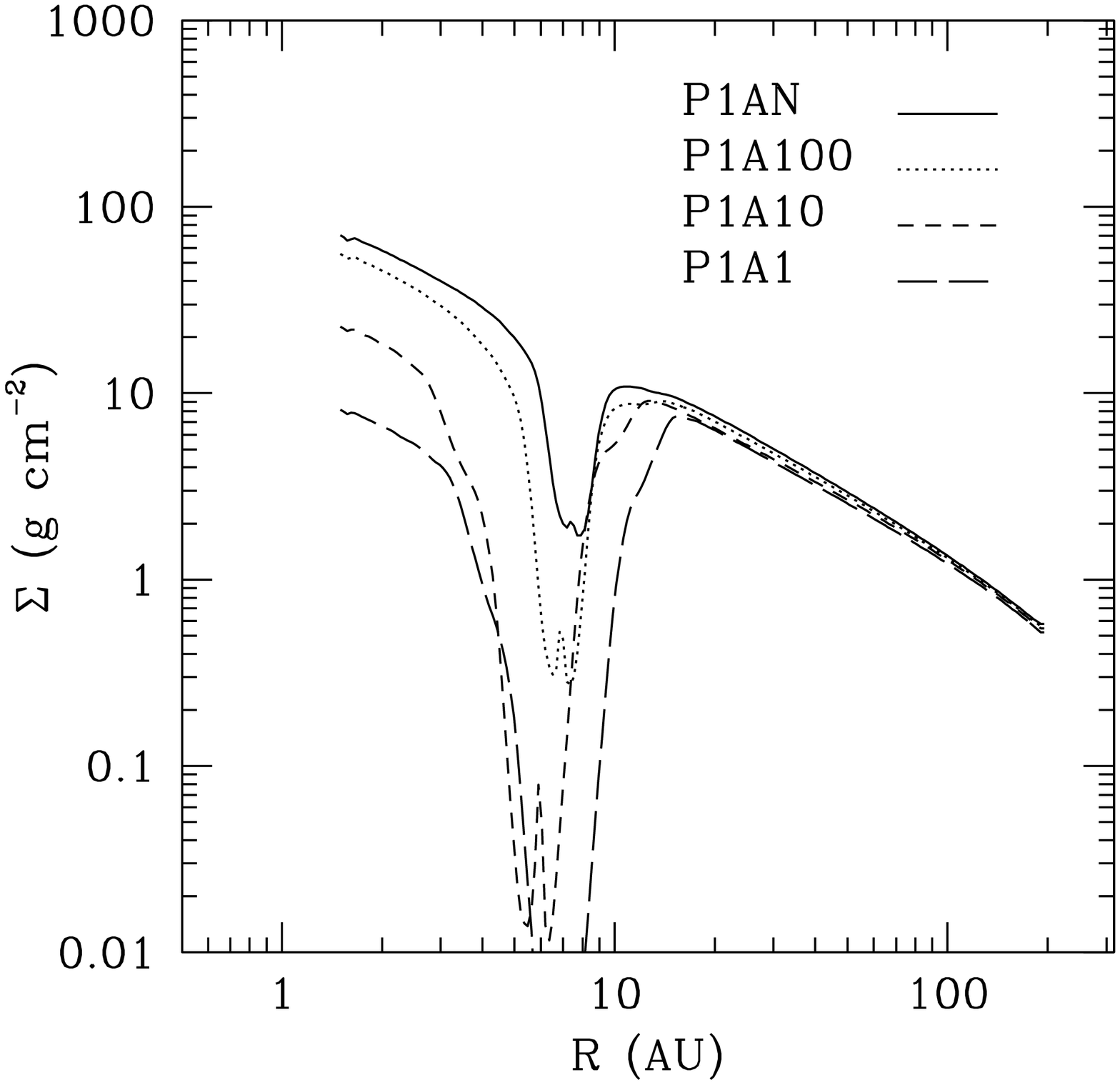} \hfil
\includegraphics[width=0.42\textwidth]{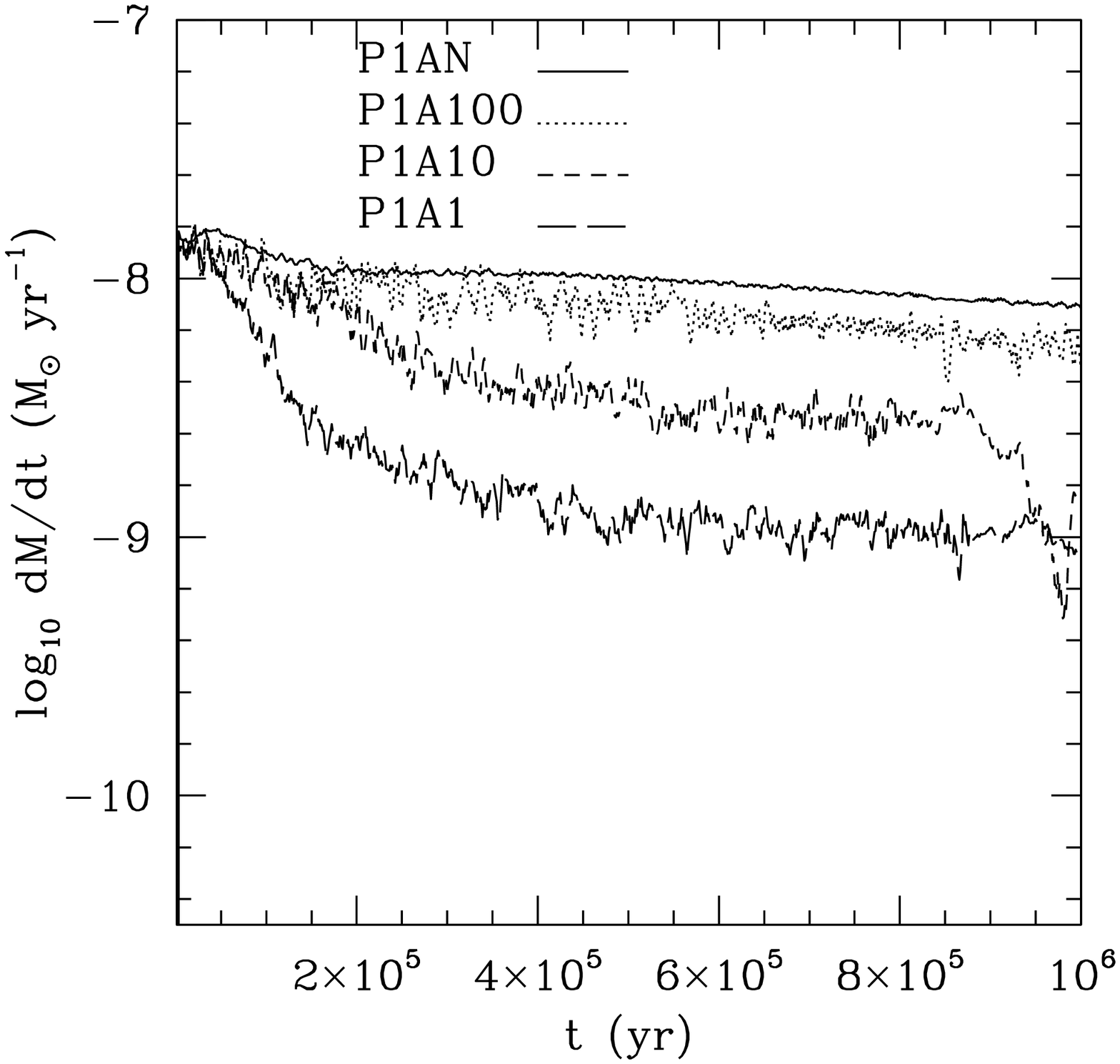} \\
\caption{Left: disk surface densities at 0.5 Myr
for viscous disks ($\alpha$=0.01)
with one planet accreting at different rates:
no planet accretion (solid curve), accretion on 100 ($f$=100) (dotted curve),
10 ($f$=10) (short-dashed curve), and
1 orbital timescales ($f$=1) (long-dashed curve; see
text for further explanation).
Right: the disk accretion rates onto the central star for these four cases.
A comparison between the left and right panels demonstrates that
the depletion of the inner disk surface density results in a proportional
decrease in accretion rate onto the central star.}
\label{fig:fig1p}
\end{figure}

We then placed a single $1 M_J$ (ramping up to $1 M_J$ during the first
5000 yrs as discussed in \S 3), non-accreting planet
in the disk at 20 AU initially (case P1AN).  After 0.5 Myr the planet
has migrated in to $\sim 8$~AU (Fig. \ref{fig:fig1p}).
The azimuthally-averaged surface density within the gap
is reduced by about an order of magnitude
(solid curve in the left panel of Fig. \ref{fig:fig1p}).
The resulting gap is relatively shallow, as expected due to the high
disk viscosity (Lin \& Papaloizou 1993; Crida \& Morbidelli 2007) and relatively large
disk thickness at 20 AU where $H/R \simeq 0.061$.
The viscous criterion for gap formation is $q > 40/R_e$,
where $q$ is the planet-star mass ratio and $R_e$ is the
Reynolds number. At 20 AU in a disk with $\alpha=0.01$
clear gap formation is only expected for a planet with
mass $M_P > 1.5 M_J$. The thermal condition for gap formation,
which determines if the disk response to the planet is non linear,
requires $R_H > H$ at the planet location, which is only just
satisfied for a $1 M_J$ planet at 20 AU.

Because the planet is not allowed to accrete from the disk,
and the tidal torques that it exerts are not strong enough
to truncate the mass flow through the disk,
accretion flow is channeled through the gap
and arrives in the inner disk. Due to the continuous replenishment
from the outer disk beyond the planet, the inner disk can maintain the $10^{-8} \msunyr$
accretion rate onto the central star (right panel of Fig. \ref{fig:fig1p}).
As expected, the inner disk surface density is not strongly affected by
the presence of the planet and is close to that obtained in the unperturbed viscous disk
model PN described above.

We next allowed the planet to accrete, at a rate controlled by
the parameter $f$ as described in \S 2.
The planet starts out with a mass of $0.1 M_J$  and is again placed at 20 AU.
If $f = 1$ (case P1A1), after the initial phase when the
planet accretes material around it and opens a gap,
the inner disk surface density decreases by one order of magnitude, and
consequently the accretion rate onto the central star
also decreases by one order of magnitude (10$^{-9}\msunyr$) relative to the
non-accreting case.  This case is nearly equivalent to that studied by
Lubow \& D'Angelo (2006) \footnote{Although their $f =0.1$,
the radius within which mass is added to the planet is 0.2 Hill radii.},
who similarly found that $\sim 90\%$ of the disk accretion flows onto the planet,
leaving only $\sim$ 10\% to pass through into the inner disk.
On the other hand we have much higher
disk surface densities inside the gap than do Varniere \etal (2006).
As discussed in \S 2,
this difference is almost certainly due to the difference in the inner boundary condition.
In test calculations performed which did not use the $\beta$-limiter on the radial velocity (\S 2),
we observed a similar rapid depletion of the inner disk (see also Crida \& Morbidelli 2007).

\begin{figure}
\includegraphics[width=0.42\textwidth]{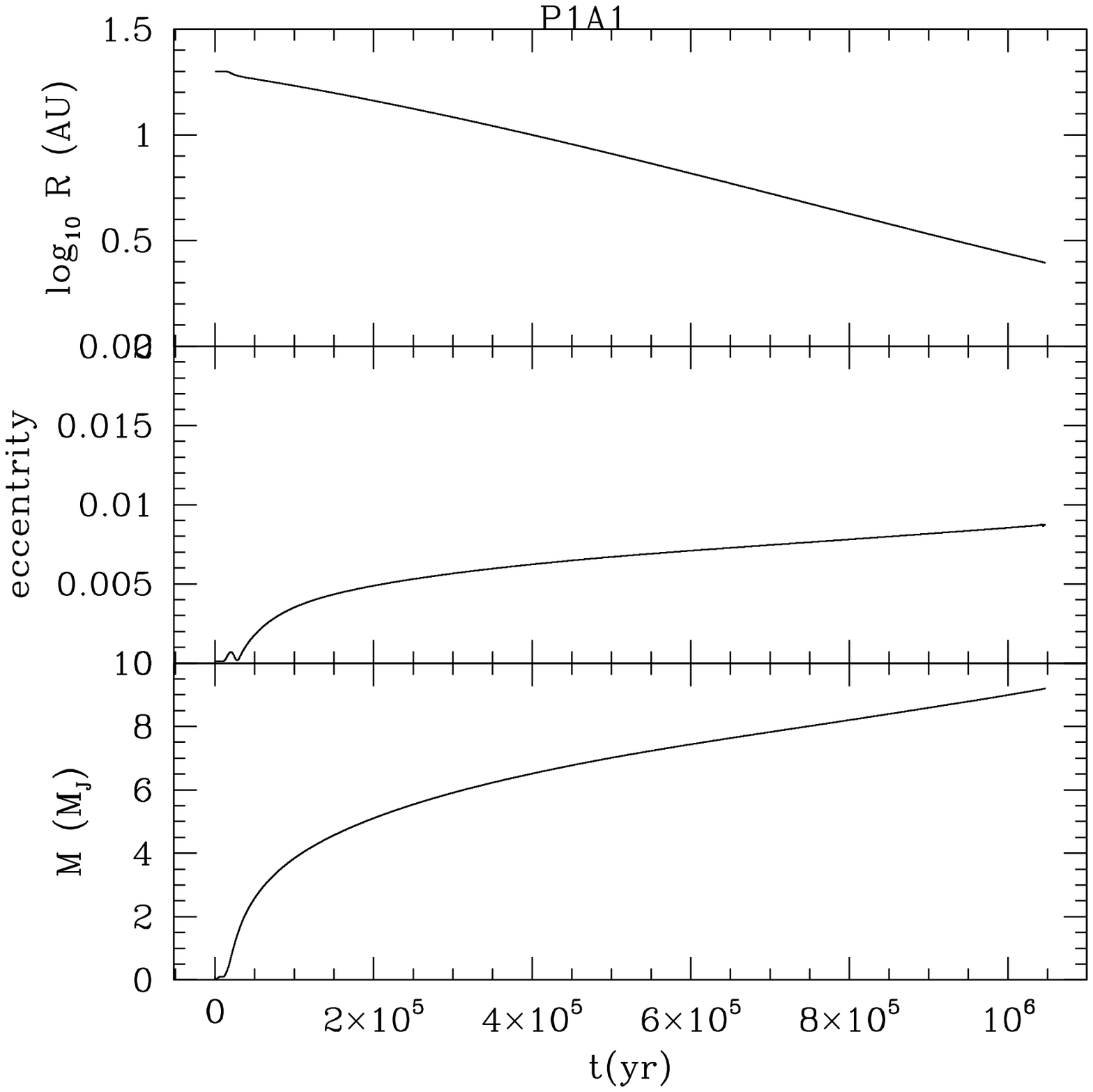} \hfil
\includegraphics[width=0.42\textwidth]{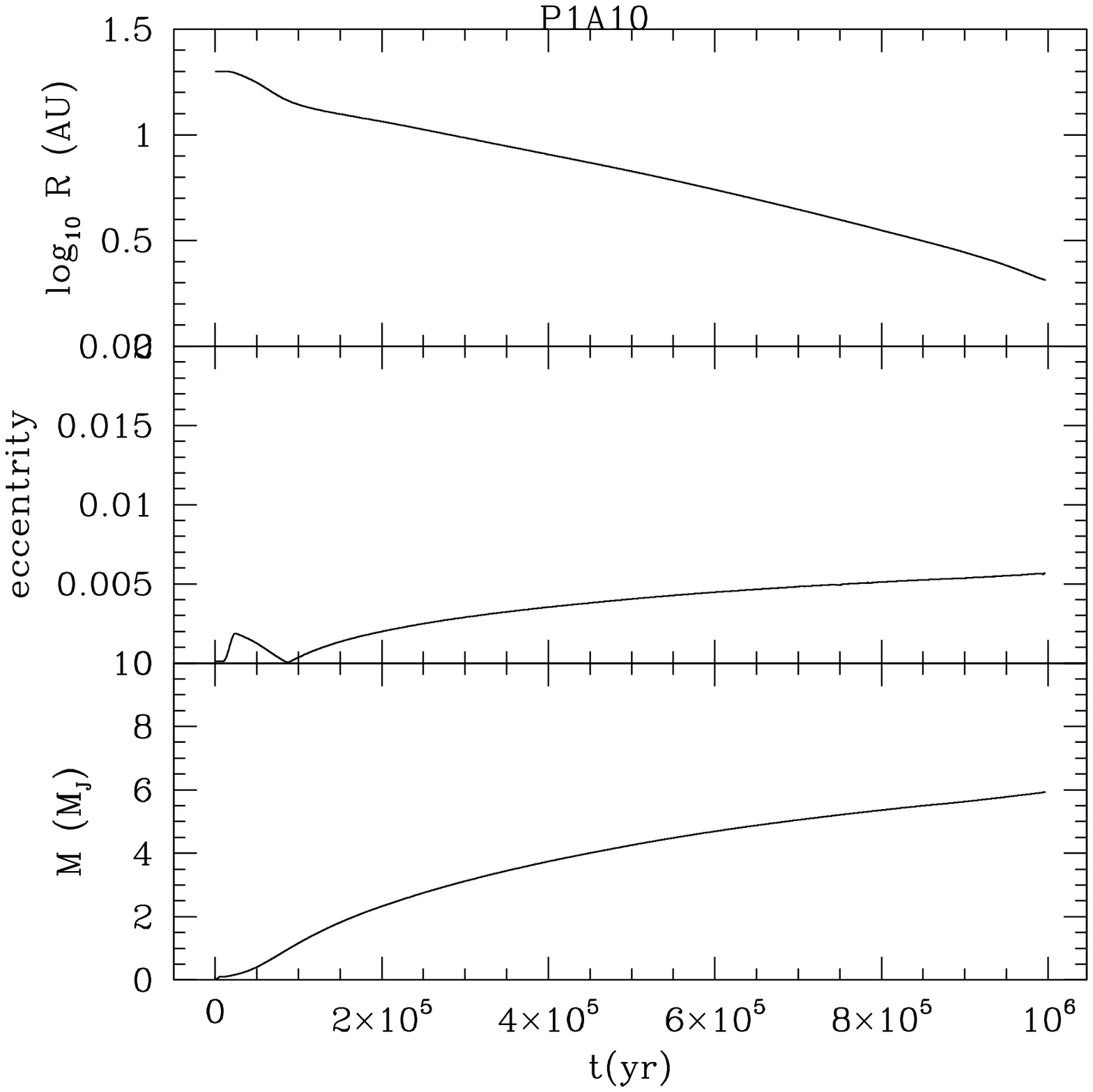} \\
 \caption{The planet's semi-major axis, eccentricity
, and mass
as a function of time for the P1A1 case (left) and P1A10 case (right).} \label{fig:AM2P1A1}
\end{figure}

As the planet mass grows with time, the gap becomes wider and deeper due to the
removal of gas from the disk and the increasing effectiveness of the tidal torques.
As 90\% of the accreting disk mass is added to the planet
at a rate $\sim 10 ^{-8}\msunyr$, the planet grows to $10 M_{J}$ after 1 Myr
(left panel in Fig. \ref{fig:AM2P1A1}),
so that the gap is much deeper than the case where the planet does
not accrete (left panel in Fig. \ref{fig:fig1p}).

If $f = 10$ (case P1A10),
the inner disk surface density and mass accretion rate
onto the central star is reduced to 30\% of the non-accreting planet case.
Because the planet accretes more slowly than in the  $f=1$ case, the planet
is less massive (right panel in Fig. \ref{fig:AM2P1A1}) and
so the gap is shallower. The eccentricity for this case
is lower than for P1A1 case (this is also true for multiple planet cases,
as discussed below).

If $f = 100$ (case P1A100), the inner disk surface density and accretion rate
onto the central star is barely affected by the planet
(dotted curve in Fig. \ref{fig:fig1p}).
Only 10\% of the mass flowing through the gap accretes onto the planet.
Thus, it takes 1 Myr for the planet to grow to $1 M_J$.
The accretion rates show an initial rapid decay due to the inner disk adjusting
to the reduced flow from the outer disk; thereafter, the accretion rates show a slow
decay or nearly-constant behavior, depending upon how fast the
planet mass grows (which affects the mass transfer through the gap).
With $\alpha$ as large as 0.01, the gap is shallow and the planet is
not fully locked to the viscous evolution of outer edge of the gap,
as there remains significant mass flow through the gap.
The resulting planet migration timescale is
$\sim 10^{6}$~yr (Fig. \ref{fig:AM2P1A1})
while the viscous timescale at that radius is $1.8 \times 10^{5}$~yr; thus
the planet does not migrate very far during the viscous timescale, allowing the
inner disk mass accretion rate to approach a steady state. The reason for
the migration being so much slower than the viscous evolution rate
is that the flow of gas through the gap causes a positive corotation torque
to be exerted on the planet (Masset 2002), reducing the rate of inward migration.
This timescale estimate also suggests that if
$\alpha$ is as large as 0.01, the supply of gas
to accrete onto the star in transitional and pre-transitional disks must come from the outer
disk to be sustained for $\sim 1$~Myr.

\subsection{Viscous disks with two planets}

\begin{figure}
\includegraphics[width=0.42\textwidth]{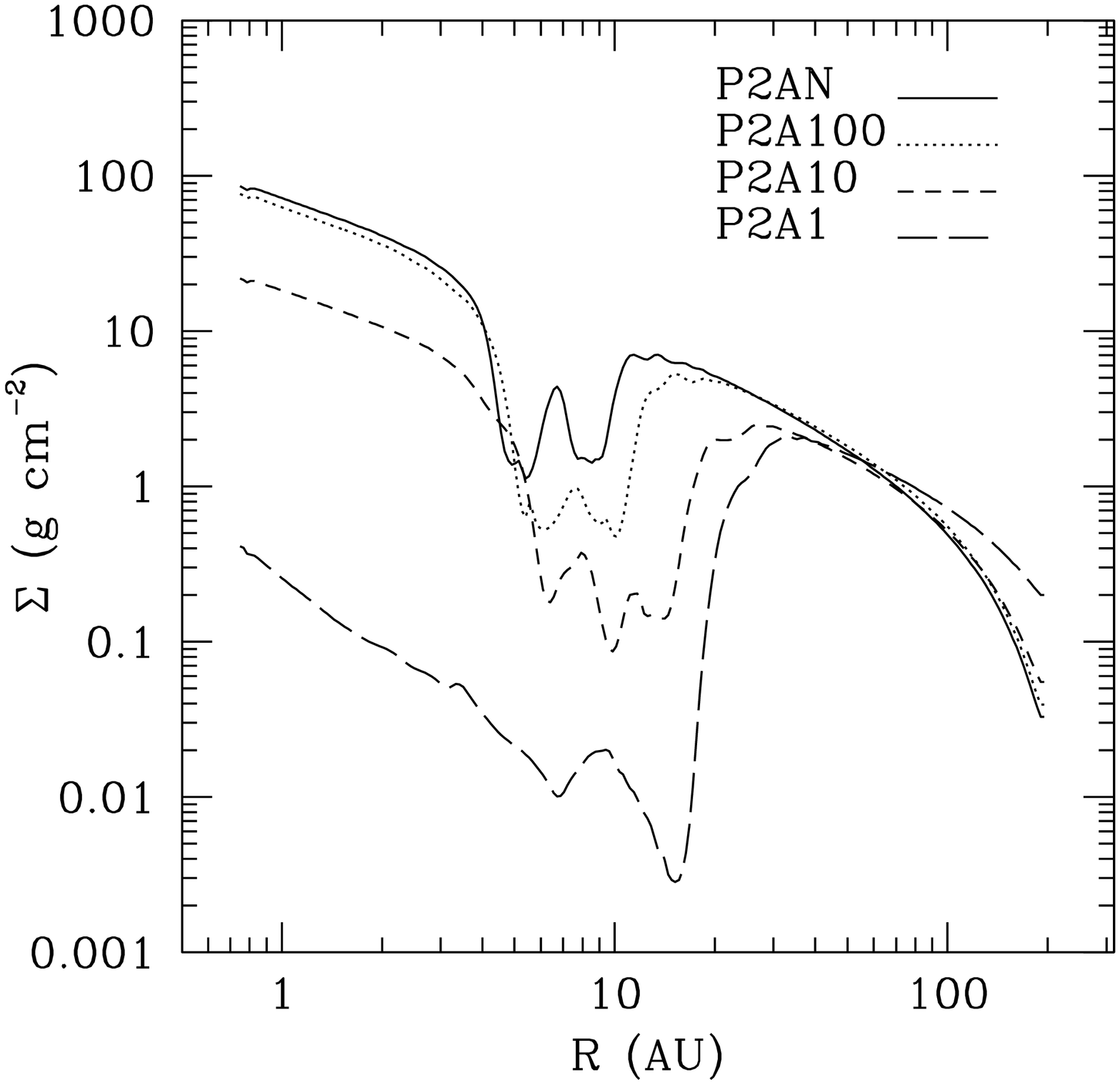} \hfil
\includegraphics[width=0.42\textwidth]{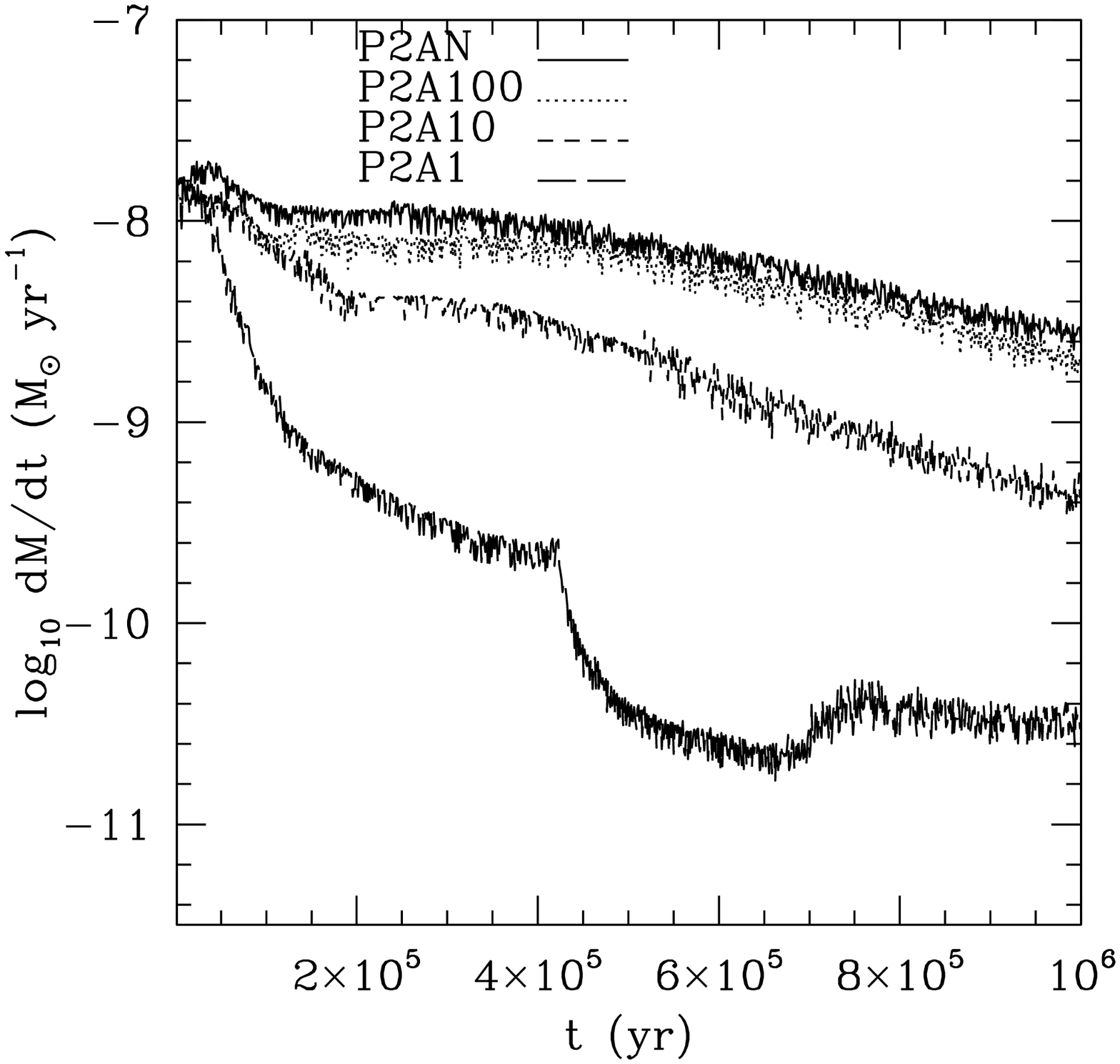} \\
\caption{Left: disk surface densities at 0.5 Myr
for two planets with no planet accretion (solid curve), $f$=100 (dotted curve),
$f$=10 (short-dashed curve), and $f$=1 (long-dashed curve).
Right: the disk accretion rates onto the central star for these four cases.}
\label{fig:fig2p}
\end{figure}

Figure \ref{fig:fig2p} shows the cases (P2A) with two
accreting planets originally placed at 12.5 and 20 AU.
The gap opened by the joint action of the two planets is considerably
wider than in the single planet case (by roughly a factor of three between the
outer and inner edge of the gap).

If neither of the planets are accreting (P2AN),
the inner disk surface density (solid curve in Fig. \ref{fig:fig2p})
and the mass accretion rate onto the star are close to the no planet case as expected after
a period of initial relaxation.
However, if both of the planets are accreting rapidly ($f=1$, P2A1)
the accretion rate onto the central star
and the surface density in the disk interior to the gap
are both reduced by almost two orders of magnitude
(long-dashed curve in Fig. \ref{fig:fig2p}),
as expected since each planet accretes roughly
90\% of the viscous flow, leaving only 1\% to make it through the gap.

\begin{figure}
\includegraphics[width=0.42\textwidth]{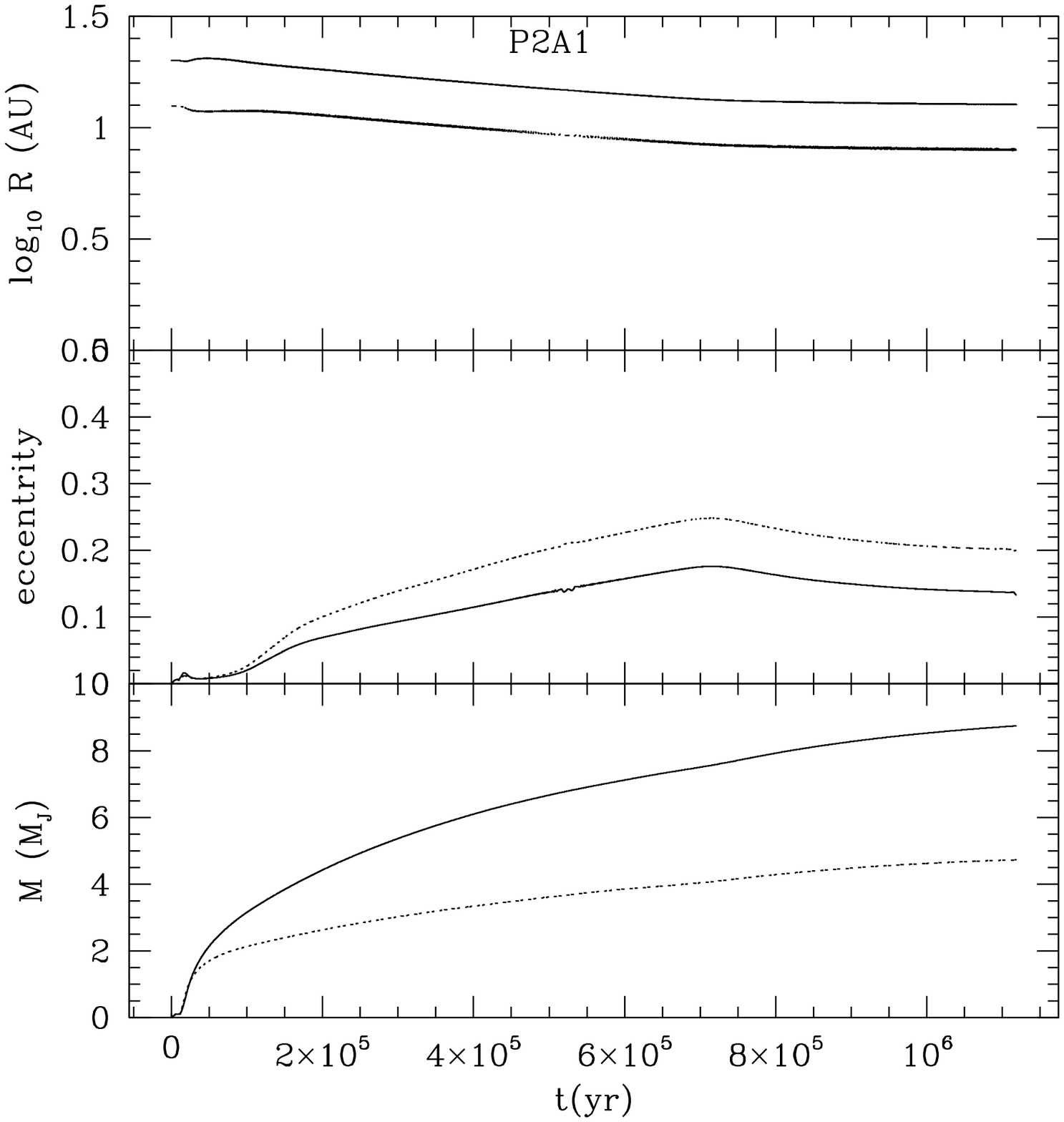} \hfil
\includegraphics[width=0.42\textwidth]{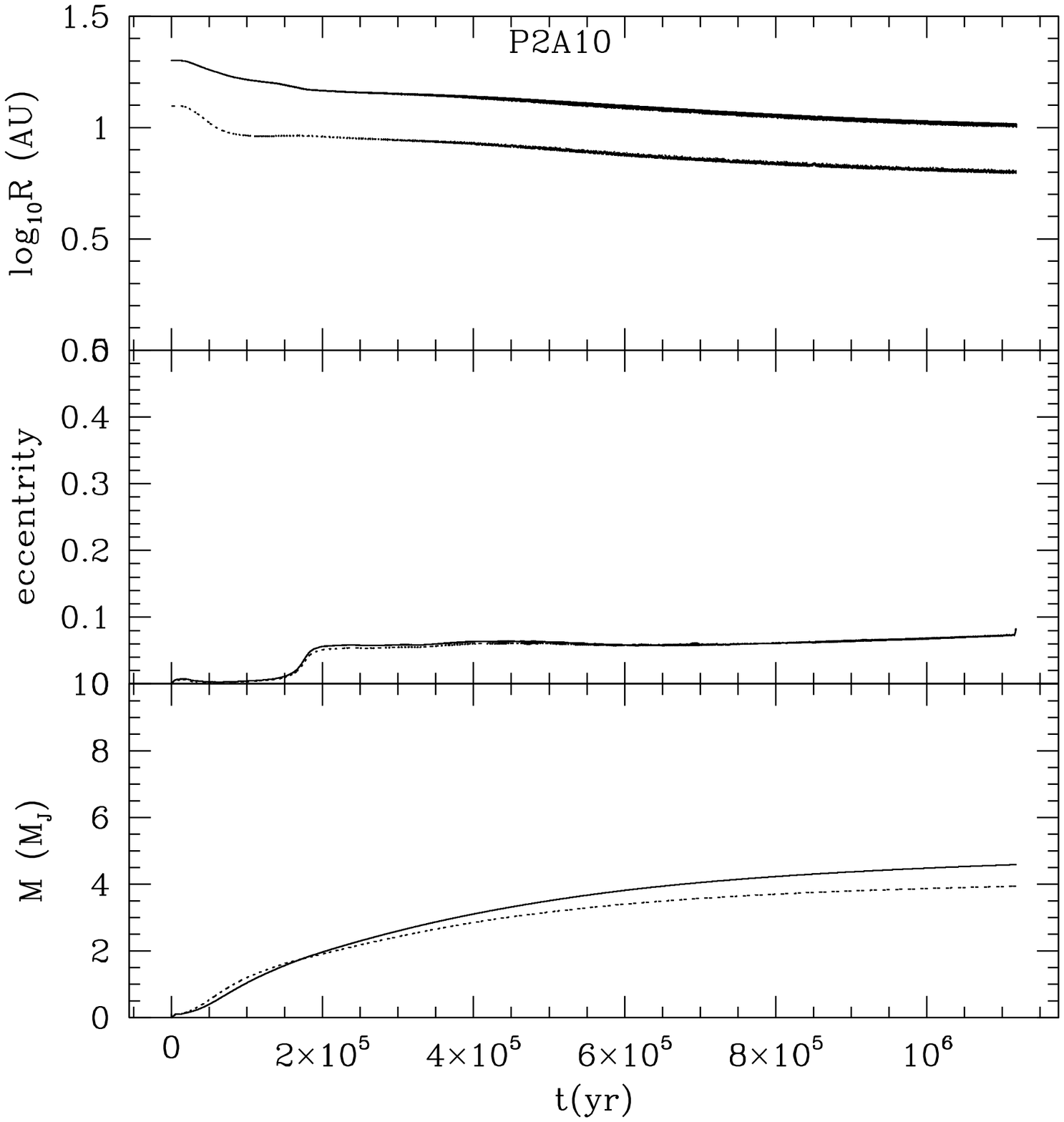} \\
\caption{The planets'
semi-major axis, eccentricities, and
masses as a function of time for the two accreting planets in the P2A1 case (left) and P2A10 case (right). The solid curves represent the outer planet while the dotted curves represent the inner planet.} \label{fig:AM2P2A1}
\end{figure}

The planets' orbital properties for the
P2A1 case are shown in the left panel of Figure \ref{fig:AM2P2A1}.
The two planets migrate inwards at a significantly
slower rate than in the one planet case (Kley 2000; compare Fig. \ref{fig:AM2P1A1}).
Due to the common gap formed by two planets, the torque from the
outer disk on the inner planet is almost zero.
Although the outer planet tries to migrate toward the inner planet, the
latter pushes the outer planet outwards by locking it into 2:1 resonance.
The eccentricity of the planets is also driven up as they move into 2:1
resonance (Fig. \ref{fig:AM2P2A1}), as seen in other
simulations (e.g., Snellgrove et al. 2001; Kley et al. 2005; Pierens \& Nelson 2008).
The outer planet's mass grows faster than the inner planet because
the outer planet is continuously being fed by material from the outer disk.

With median and slow planet accretion rates ($f$=10 and 100, cases P2A10, P2A100),
the disk accretion rates onto the central star and the inner disk
surface densities are reduced by factors similar to that of the
equivalent single planet cases. Unlike the P2A1 case,
both of these planets almost grow at the same rate
(right panel of Fig. \ref{fig:AM2P2A1}) because
there is enough mass flow passing the outer planet to feed the inner planet.

\subsection{Viscous disks with three and four planets}
\begin{figure}
\includegraphics[width=0.42\textwidth]{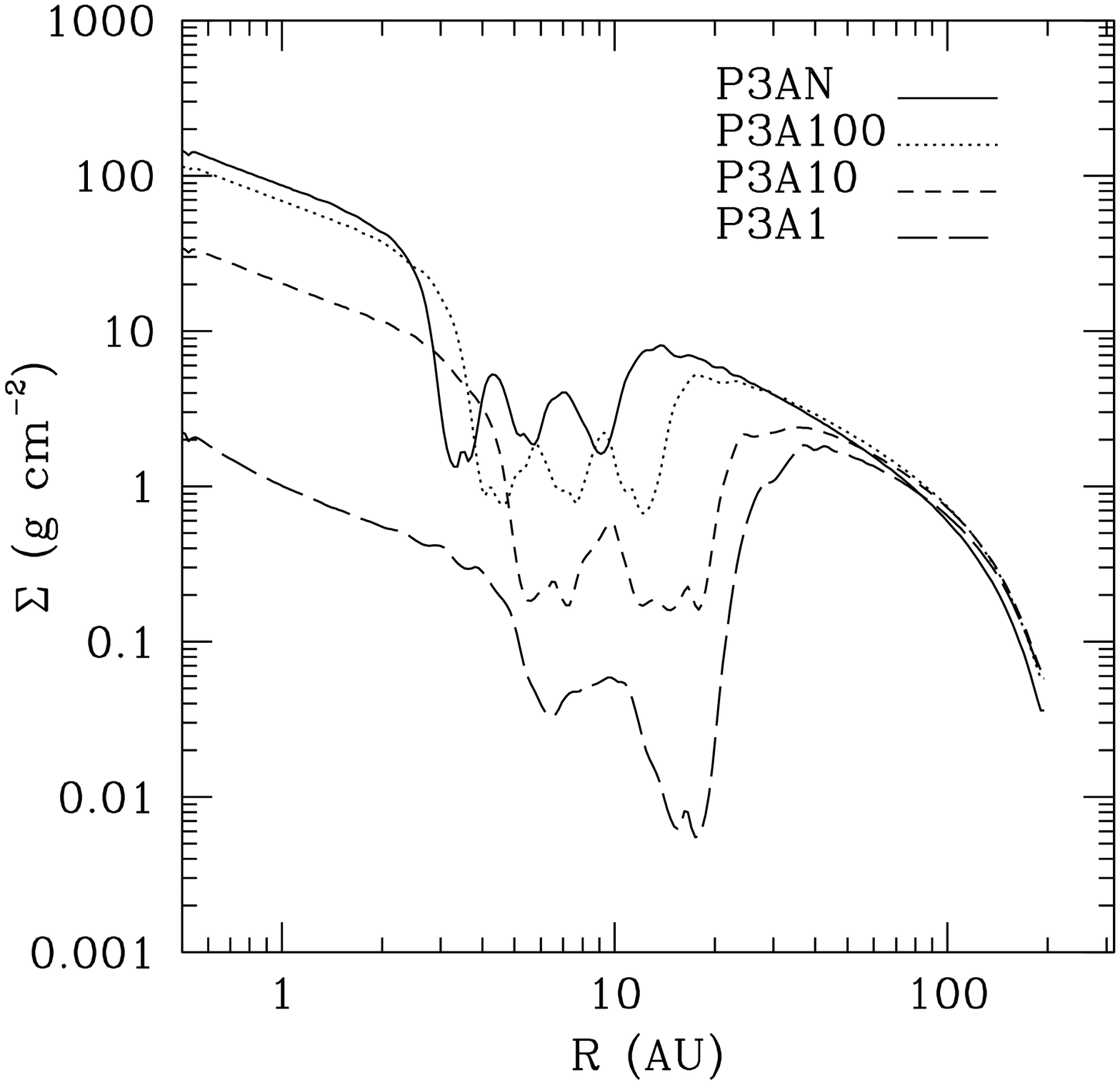} \hfil
\includegraphics[width=0.42\textwidth]{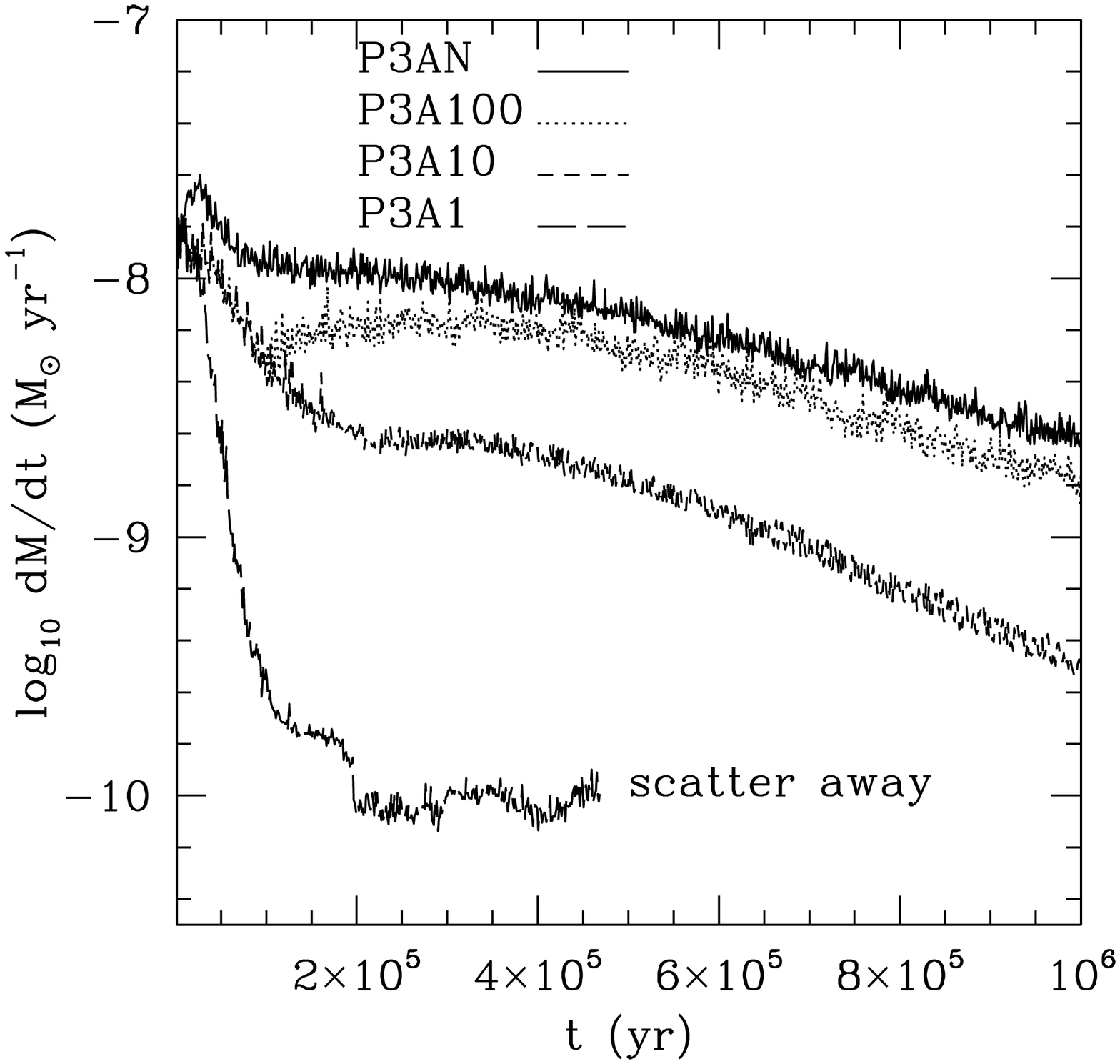} \\
\caption{Left: disk surface densities at 0.4 Myrs
for three planets with no planet accretion (solid curve), $f$=100 (dotted curve),
$f =10$ (short-dashed curve), and $f=1$ (long-dashed curve).
Right: the corresponding disk accretion rates.}
\label{fig:fig3p}
\end{figure}

Including more giant planets in the disk produces wider gaps, as expected
(see Figs. \ref{fig:fig3p} and \ref{fig:fig4p} for the case with
three and four planets, respectively).
However, the greater the number of planets that are present, the more the
accretion flow reaching the inner disk is reduced.
Fast planetary accretion ($f$=1) quickly results in decreasing
both the inner disk surface density and the mass accretion
rate by more than two orders of magnitude, while median and slow planet accretion
($f$=10 and 100) lead to less than one order of magnitude reduction initially.
There is a longer-term decay of the mass accretion rate,
which is due to the fact that the inclusion of more planets disturbs the disk more,
and over long time scales this causes mass to be depleted from the disk
as it escapes through the open outer boundary.

\begin{figure}
\includegraphics[width=0.42\textwidth]{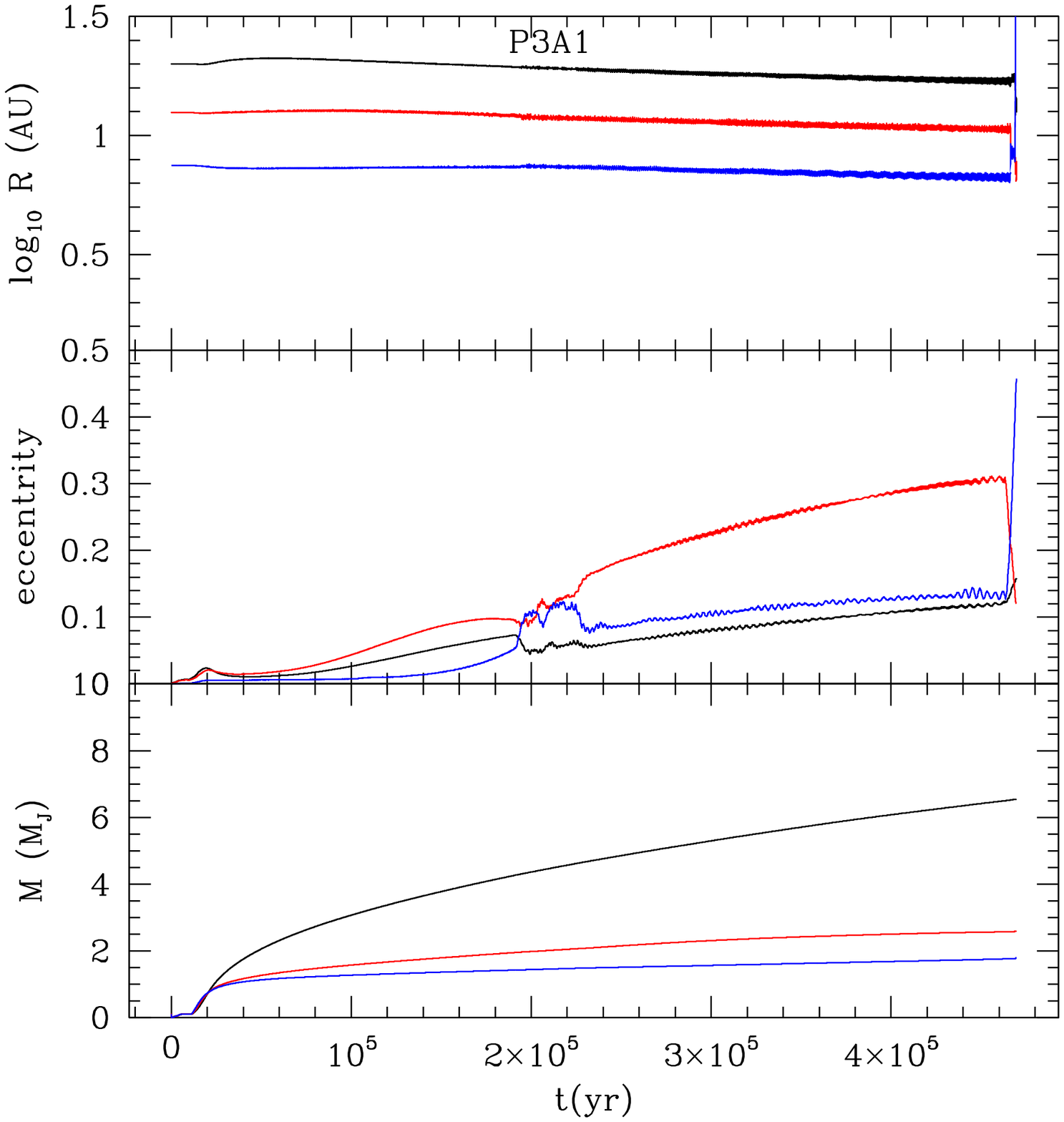} \hfil
\includegraphics[width=0.42\textwidth]{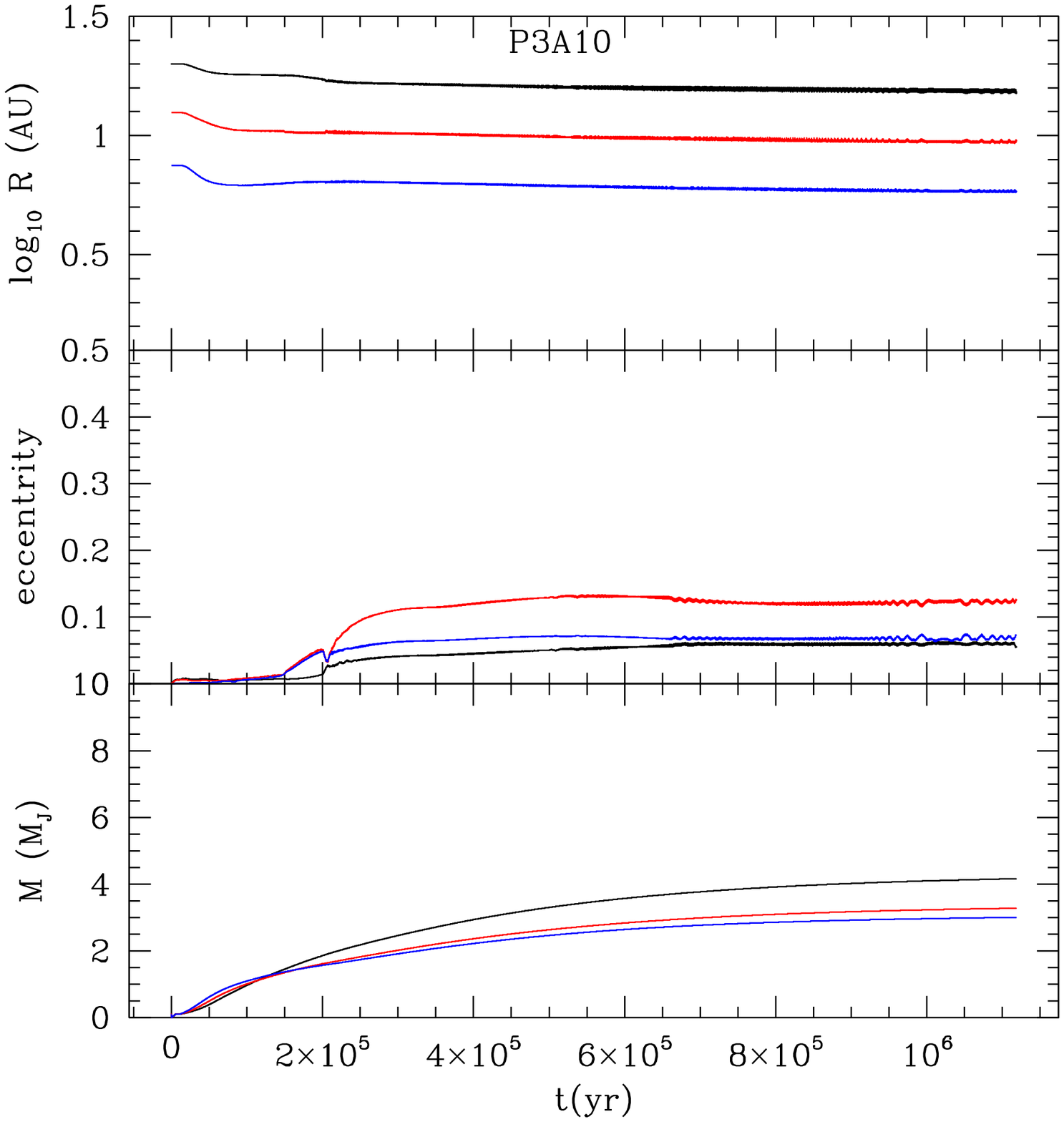}\\
\caption{The planets'
semi-major axis, eccentricities, and
masses as a function of time for three accreting planets in P3A1 case (left) and P3A10 case (right). } \label{fig:AM2P3A1}
\end{figure}

\begin{figure}
\includegraphics[width=0.42\textwidth]{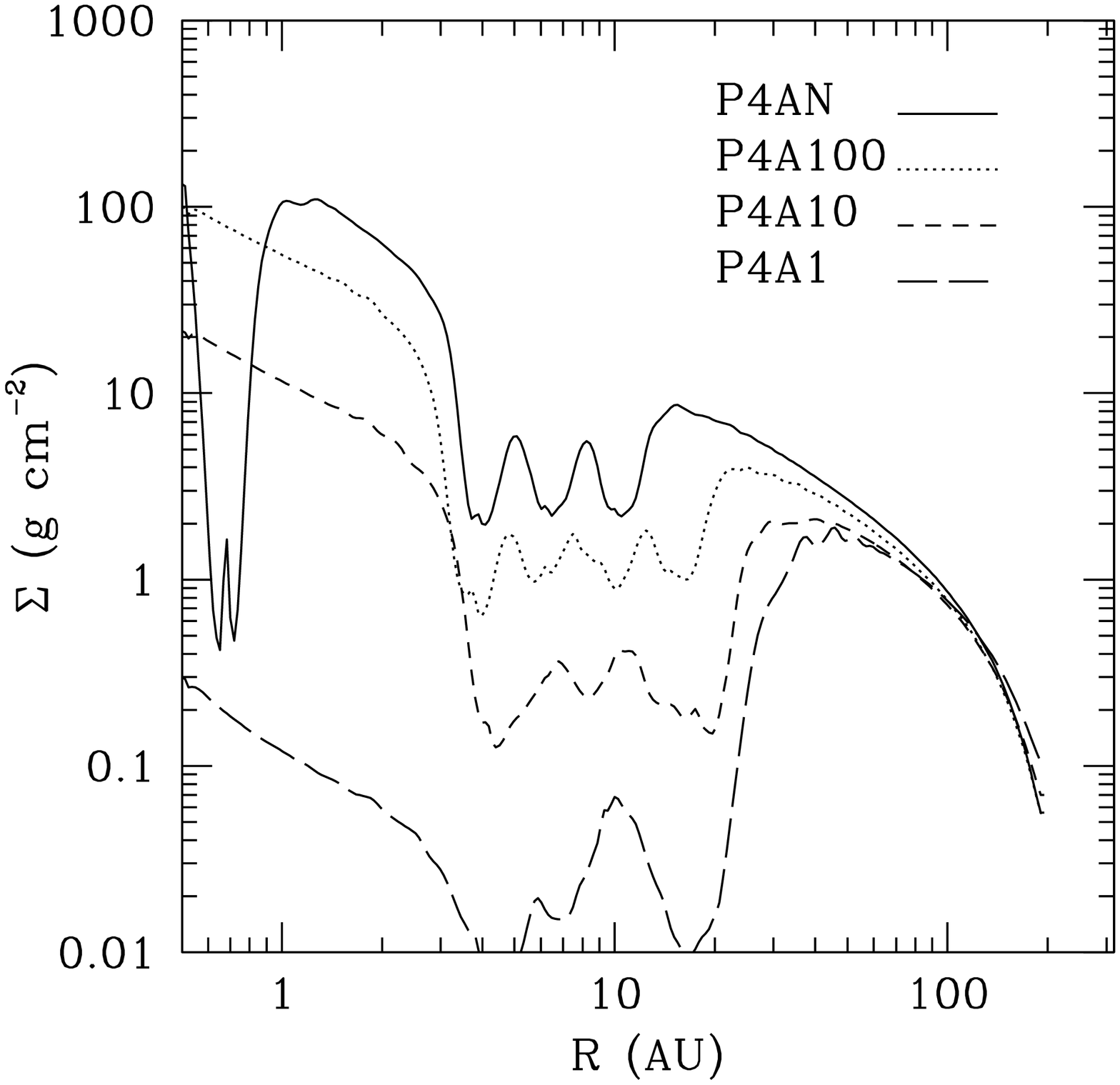} \hfil
\includegraphics[width=0.42\textwidth]{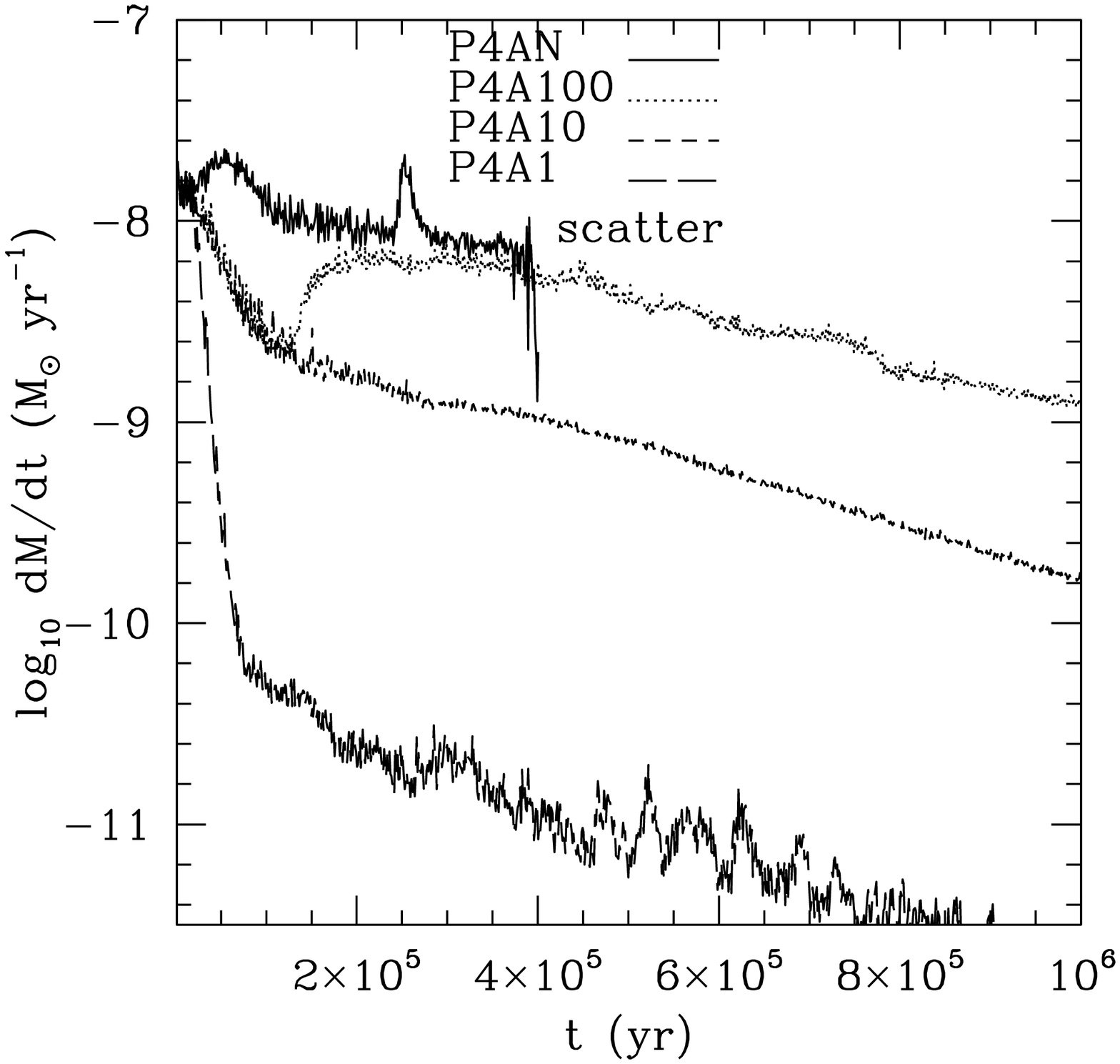} \\
\caption{Left: disk surface densities at 0.4 Myrs
for four planets with no planet accretion (solid curve), $f$=100 (dotted curve),
$f =10$ (short-dashed curve), and $f=1$ (long-dashed curve).
Right: the corresponding disk accretion rates.}
\label{fig:fig4p}
\end{figure}

\begin{figure}
\includegraphics[width=0.42\textwidth]{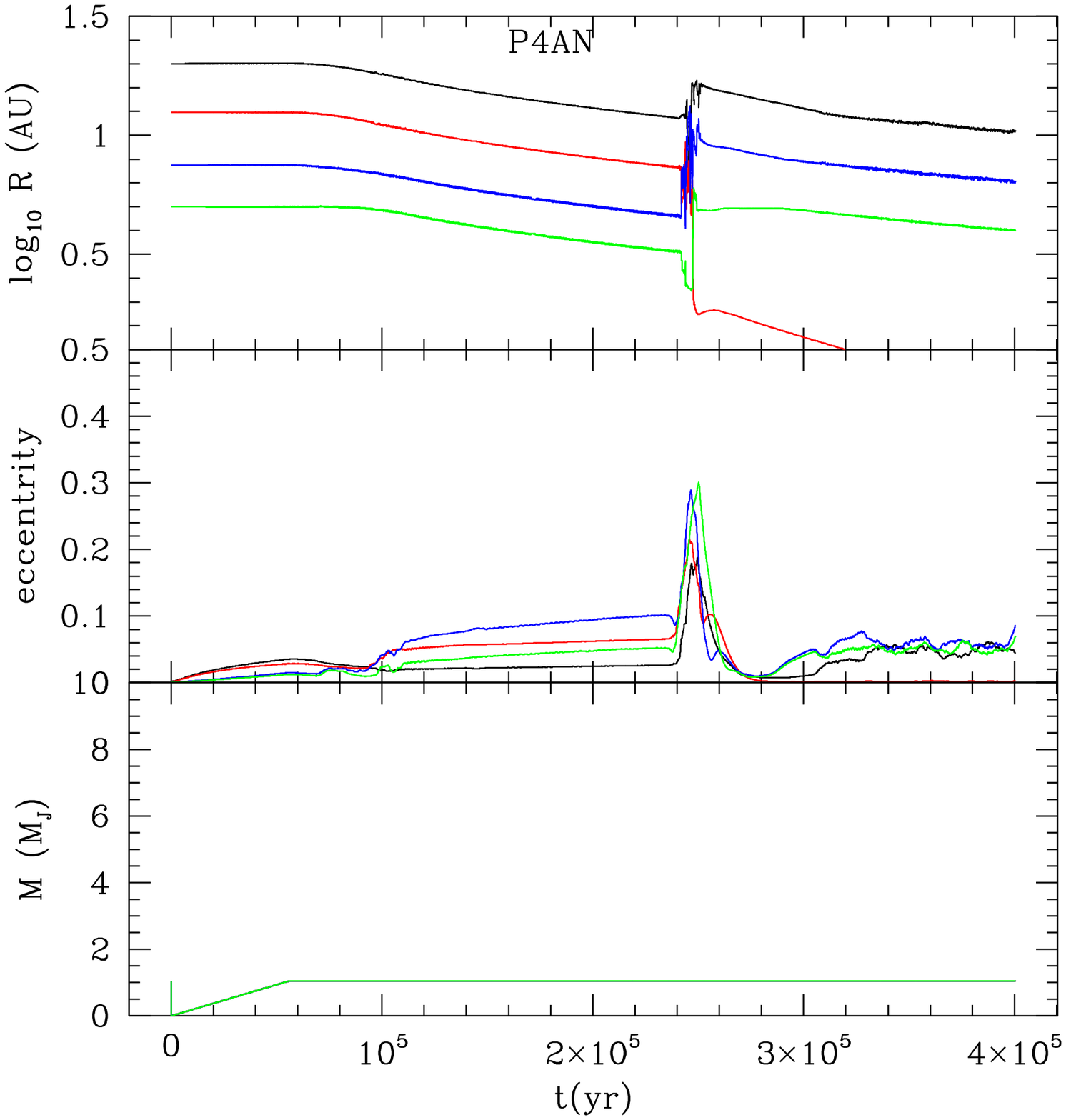} \hfil
\includegraphics[width=0.42\textwidth]{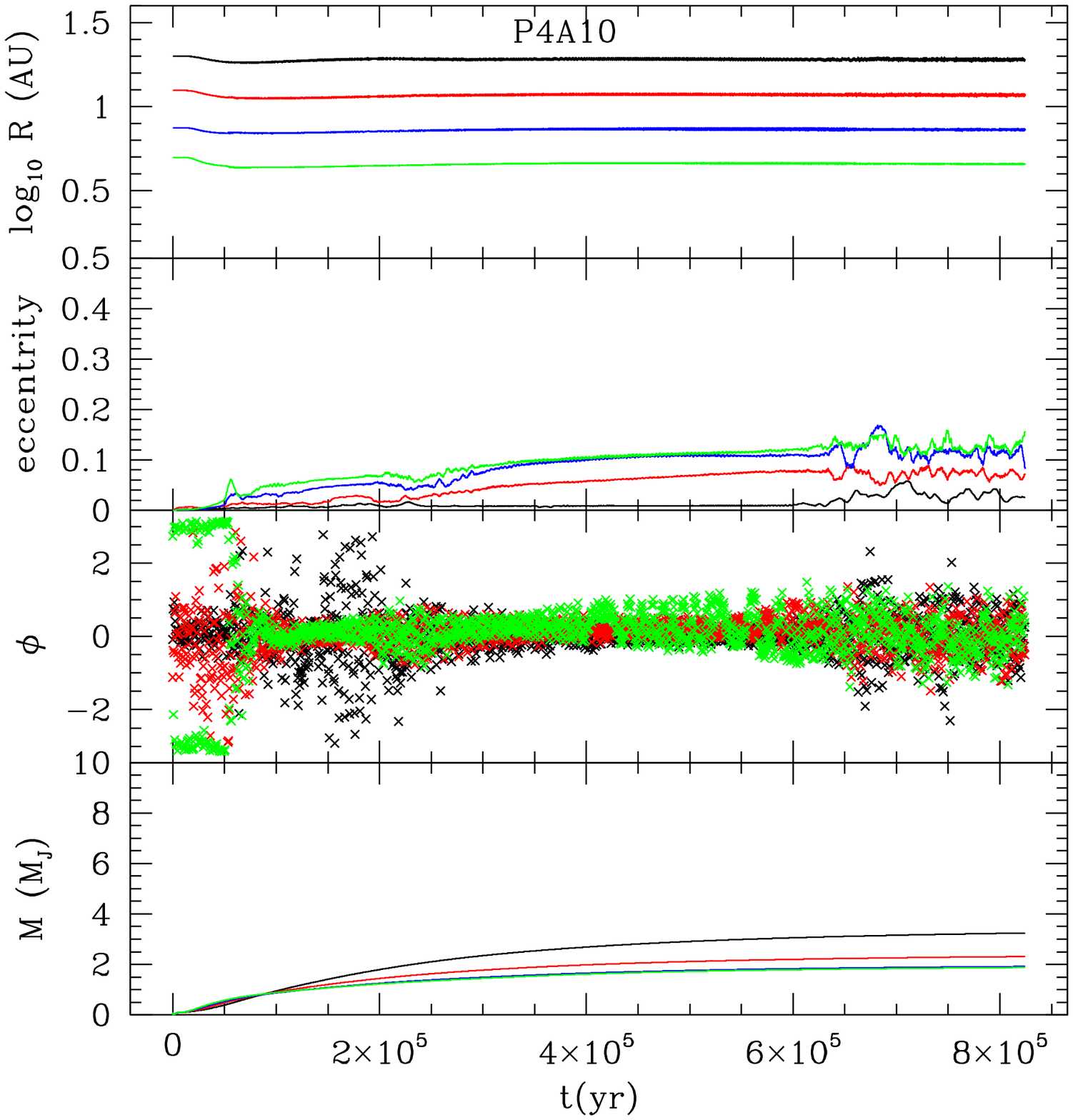}\\
\caption{The planets' semi-major axis, eccentricities, resonant angles (2:1),
masses with time for four planets in P4AN case (left) and P4A10 case (right).
In the $\phi$ panel of the right figure, black, red, and green curves represent
the resonant angle between the pairs of 4/3, 3/2, 2/1 planets
(4 is the outermost one.).} \label{fig:AM2P4A10}
\end{figure}

The planets' orbital properties are shown in
Figs. \ref{fig:AM2P3A1} and \ref{fig:AM2P4A10}.
In most cases, the planets are trapped into 2:1 resonances between each
pair of adjacent planets. The middle panel of Fig. \ref{fig:AM2P4A10}
displays the time evolution of the resonant angle
$\phi$=2$\lambda_{O}$-$\lambda_{I}$-$\varpi_{I}$ associated with the
2:1 resonance, where $\lambda_{I}(\lambda_{O})$ and $\varpi_{I}(\varpi_{O})$
are respectively the mean longitude and longitude of pericenter of the
inner (outer) planet of the pair. Whenever the resonant angle librates between
the interval [-$\pi$,$\pi$] the planets are trapped in a 2:1 resonance.
The smaller the amplitude of libration, the deeper resonance locking. As in the
two planet case (Figure \ref{fig:AM2P2A1}),
the planets' migration slows down due to resonance locking, possibly
as a result of the growth of planetary eccentricity which can reduce the migration
torques, and also due to the larger inertia associated with a resonant
multi-planet system.

There are two cases, however, where planets experience a
close encounter and are gravitationally scattered.
The first case is P3A1 (left panel in Fig. \ref{fig:AM2P3A1}),
where the eccentricity of the middle planet is driven so high that
its orbit overlaps with the inner planet's orbit, leading the inner planet
to be scattered out from this system (in a hyperbolic orbit).
The second case is P4AN (left panel in Fig. \ref{fig:AM2P4A10}),
where the innermost two planets slip out of 2:1 resonance
leading to scattering. In this case the planets remain
bound but switch their positions. It is clear that the disk model
with $\alpha=0.01$ leads to circumstances where three or four planet
resonances can be maintained stably over Myr time scales, but
can also lead to the break up of the resonances. This diversity of
behavior appears to have a strong stochastic component, such
that the stability of the planet system depends on the detailed
history of disc-planet and planet-planet interactions.

We also performed some limited calculations of
three or four planets in disks with a lower viscosity
($\alpha$=0.002) and a higher disk mass to give the same disk accretion rate.
In contrast to the lower mass $\alpha=0.01$ disks, these models
all produced unstable planet systems which resulted in scattering
and ejection of at least one planet from the system.
This apparently occurs for two reasons. The first is that the
low $\alpha$ disks result in deeper gaps, leading to a reduction
in the eccentricity damping provided by the disk, thus favoring planetary
instability. Secondly, the significantly larger disk mass probably has a more
important dynamical influence on the resonant planetary systems,
possibly helping to render the resonant configuration unstable.
We will consider the effect of varying the viscosity in more detail in
a future paper.

\section{Discussion}

Returning to the questions posed at the end of the Introduction, our findings
are as follows:

1) As seen in previous simulations, a single planet opens up a small gap.
Multiple planets can open wider common gaps.
However, in order to explain the pre-transitional disk gaps
spanning almost an order of magnitude  in radius
without dust depletion (\S 2),
we need as many as three to four giant planets,
given the tendency of the viscously evolving disk to drive the planets
into 2:1 resonances. This is a large number of such planets, given current exoplanet
statistics which do not include examples of systems containing more
than two planets in resonance. One possibility is that resonant planet systems
which contain more than two planets are able to remain stable over significant
time periods ($\sim 1$ Myr) in the presence of the gas disk, whose contribution
to eccentricity damping helps maintain dynamical stability. Once the gas disk
dissipates, however, these resonant planetary systems may become unstable,
leading eventually to ejection of some members and the formation of more
sparsely populated systems of planets on eccentric orbits, similar to those observed.
According to this hypothesis, large systems of planets help explain the presence of
the large disk gaps in pre-transitional disks, and also explain the population of
eccentric giant exoplanets.

In principle, the gaps could be widened for fewer planets if the orbits were eccentric;
however, with large $\alpha$=0.01,
we do not see high eccentricities in stable systems.

2) A high accretion rate past the planetary gap can be maintained
if accretion onto the planets is sufficiently slow.  Planetary accretion rates
of $f=10$ still permit substantial accretion
($>$10\% accretion rate outside the gap) past the
planets to the inner disk.  However, the reduction in surface density
interior to the gap is in proportion to the reduction in mass accretion
rate.  This differs from the results of Varniere \etal (2006) but is
consistent with the results of Crida \etal (2007), and is almost certainly
the result of the improved treatment of the inner boundary condition.

3) Carefully comparing our simulations with observations indicates that the pre/transitional disk systems
require substantial reduction in dust opacities within the gaps (and inner disk
if present) since the reductions in surface density there are not sufficient to
explain the very small amount of dust required to fit the observations.
For example, if we adopt the ISM opacity at 10 $\mu$m
(the assumption used by Espaillat \etal (2010) to estimate optically-thin dust masses)
of $\kappa \sim$ 10 cm$^{2}$ g$^{-1}$,
even the gaps with surface densities $\Sigma \gtrsim 0.1 {\rm g\, cm^{-2}}$ are still optically
thick, which is the case for all but the $f=1$ simulations. $f=1$ simulations still have optical depth $\sim$0.1 and they are the cases with an extreme upper limit
on the expected planetary accretion rate, which leads to low accretion rate onto the star (esp. with multiple planets).
Furthermore, the dust depletion
must be larger in the inner disk than in outer regions.  This follows from
our adoption of a realistic temperature distribution and improved
inner boundary condition, which results in
a significantly higher surface density at small radii than at large radii;
thus the inner disk will not be optically thin unless the outer disk is also
optically thin, which disagrees strongly with transitional disk observations.

One alternative
is to place planets sufficiently close to the central star that the gap essentially
extends either up to the dust destruction radius, or near to it; however, in that case it
is difficult to extend the gap all the way out to 20 - 70 AU as observed
(Table 1) when using only four planets. In principle it may be possible to
accommodate additional planets at smaller orbital radii which truncate
the inner disk to a smaller radius.
The other alternative is to allow the planets to accrete more mass,
depleting the inner disk; but then the mass accretion rate onto the central
star is reduced by unacceptably large values.

Protoplanetary disks are probably not fully viscous (e.g., Gammie 1996),
but the existence of ``dead zones'' and/or reduced viscosity regions is unlikely
to improve the situation, as they will lead to higher surface densities for
a given mass accretion rate, and may reduce the stability of systems
of more than two planets in resonance.

\subsection{Gap structure and dust depletion}
\begin{figure}
\epsscale{.80} \plotone{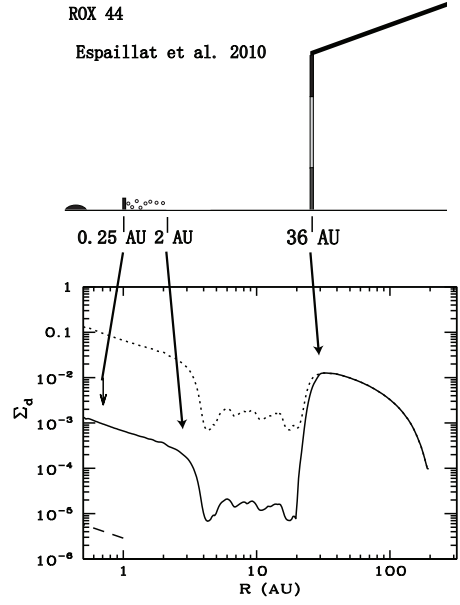}
\caption{The pre-transitional disk structure taken from Espaillat \etal (2010) compared with the azimuthal-averaged dust surface density of
the P4A10 case. The dotted curve represents the case with no dust depletion and the solid curve represents the case with the dust at the inner disk depleted by a factor of 100. The solid curve can explain the pretransitional disk structure as indicated by the arrows. The dashed curve is the dust surface density from the transitional disk GM Aur estimated by Eq. 9, which requires the dust to deplete by a factor of 10$^{5}$ by comparison with the dotted curve.} \label{fig:tau}
\end{figure}

A multi-giant planet model coupled with substantial dust depletion in
the gap and inner disk does have an attractive feature.
As pointed out in  \S 2, there is some evidence -
both from SED modeling and from near-infrared
interferometry - that the pre-transitional and transitional disks exhibit
structure within the optically-thin disk gap or hole,
such that the innermost gap region is much less cleared of small dust than
the outer gap region.  Schematically, we might identify the highly-cleared
region with the true planet-driven gap, and the inner, optically-thin dust
region with accreting gas that has been strongly depleted in dust but not so
depleted in gas surface density.  The idea is illustrated in
Fig.\ref{fig:tau}, which compares the inferred structure of
one of the pre-transitional disk systems with the
azimuthally-averaged dust surface density for the P4A10 case.

In Fig. 9,
the disk optical depth at 10 $\mu$m can be estimated by multiplying the dust surface density (dust refers to small dust particles which contribute to 10 $\mu$m opacity) by 1000 cm$^{2}$/g (which is estimated using the ISM opacity: a factor of 100 comes from the gas-to-dust mass ratio and 10 cm$^{2}$/g is the ISM opacity at 10 $\mu$m). Thus, disk regions
with dust surface density $>$10$^{-3}$ g/cm$^{2}$ are optically thick.
The dotted curve represents the dust surface density resulting from the model P4A10
assuming there is no dust depletion, which is obtained by dividing the gas
surface density from the P4A10 simulation by the nominal gas-to-dust mass
ratio $\sim$100. In this case even the gap is optically thick. The solid curve
represents the case where we notionally deplete the dust content inside
the gap by a factor of 100. This can explain the pre-transitional disks whose
inner region within $\sim 1$ AU changes from being optically thick to
optically thin, with the region beyond being almost dust free.
This comparison with pre-transitional disks set a lower limit on the dust depletion factor, while the transitional disks can set a higher limit. If we assume all the dust
inside the transitional disk gap has been detected (which is reasonable since
transitional disks are optically thin), we can estimate the dust surface density
by solving
\begin{equation}
\int_{0}^{R_{out,thin}}\Sigma_{d}(R)2\pi R dR=M_{d}\,,
\end{equation}
where R$_{out,thin}$ and M$_{d}$ are given in Table 1, and is assumed
to be $\Sigma_{d}$(AU)(R/AU)$^{-1}$. The derived $\Sigma_{d}$ for GM Aur ($<$1 AU) using Table 1 is shown as the dashed curve in
Fig. 9. By comparing with the dotted curve, we find the dust needs to
deplete by a factor of $\sim$10$^{5}$. Thus we estimate the dust-to-gas mass
ratio in the inner disk is between 10$^{-5}$ and 10$^{-2}$ of the ISM value (0.01)
for our model to be consistent with pre/transitional disks.

More broadly, a gap-opening perturbing body or bodies still remains a plausible
explanation of the pre/transitional disks.  This type of model naturally
predicts a sharp transition in surface density at the outer gap edge, which
is needed to explain the strong mid- and far-infrared emission of these systems.
While radially-dependent dust depletion might also mimic this effect,
most T Tauri disks do not show this behavior.
There are a significant number of T Tauri systems in which
the disk is optically-thick at a wide range of radii,
but which are much more geometrically flat based on their SEDs
(e.g., the ``group E'' systems shown in Figure 7 of Furlan \etal 2006).
These objects suggest that dust growth and settling can
occur roughly simultaneously over a wide range of radii, without resulting
in the abrupt change between optically-thin and optically-thick regions
that occurs in the pre/transitional disks.
In addition, the large outer disk gap radii (as much as 20-70 AU)
pose a challenge for pure coagulation models at an age of $\sim 1-2$~Myr
(Tanaka \etal 2005; Dullemond
\& Dominik 2005).

The requirement of significant dust depletion is not completely surprising,
given the evidence for small dust depletion in many T Tauri disks without
obvious gaps or holes (e.g., D'Alessio \etal 2001; Furlan \etal 2006).
In addition, formation of giant planets via core accretion clearly requires
dust growth.  Indeed, an outstanding theoretical problem has been
to avoid clearing inner disks by ages of 1 Myr or less
(e.g., Dullemond
\& Dominik 2005; Tanaka \etal 2005),
resulting in suggestions that small dust must be replenished to some extent
by fragmentation as a result of collisions of larger particles.

A hypothesis which explains many features of pre/transitional disks
using both giant planets and dust depletion can be outlined as follows.
Dust coagulation and growth in the disk interior to $\sim 20$ AU
ensues, leading to the formation of planetesimals, and eventually to the formation of
a system of numerous giant planets. This system of giant planets
forms a large common gap which covers radii from $< 1$ AU out to
$\sim 50$ AU. Remnant planetesimals within the gap region, whose collisions
can act as a secondary source of dust, are dynamically cleared out
- preventing in situ secondary dust formation. After a viscous time scale
corresponding to the size scale of the planetary system, the gas present
in the planet-induced gap and the inner disk within 1 AU has originated
largely from that part of the disk which lies out beyond the planetary system.
This gas may be substantially depleted of small dust because of significant
grain growth at large radii, combined with filtration of dust at the
outer edge of the gap (Paardekooper \& Mellema 2006; Rice \etal 2006). Furthermore,
small dust particles may manage to pass through the gap, but they can quickly coagulate
to big dust particles (Dullemond
\& Dominik 2005).
In this scenario, gas which accretes through the system at
late times is strongly depleted of dust, can sustain a significant accretion
rate onto the central star, and can provide a dust-rich wall at the outer
edge of the gap required by SED modeling. We will explore this hypothesis
in a forthcoming paper in which we include the effects of dust filtration.

\subsection{Numerical limitations}
Our simulations are highly simplified, and this needs to be borne in mind when
interpreting our results. The resolution we adopt is insufficient to
resolve the gas flow within the planet Hill sphere, and so we are forced
to adopt a simplified approach to simulating gas accretion onto the planet,
instead of simulating the accretion process directly. As such the detailed evolution
of the gas flow in and around the planet Hill sphere may not be modeled with
a high degree of fidelity in our simulations. The simulations are two dimensional,
which may not be a good approximation early on when the planet masses
are low, but improves as the planet masses become large and gap formation occurs.
We have neglected a detailed treatment of the gas thermodynamics, adopting instead
a locally isothermal equation of state. A proper treatment would allow the
local disk temperature to be determined by a balance between viscous and stellar heating,
and radiative cooling, and this might affect details of the gap structure due
to the changing optical depth of the gas and its thermal evolution there
(although viscous heating is generally much smaller than stellar heating).
Ayliffe \& Bate (2010) suggest radiation due to circumplanetary disk accretion
tends to suppress the spiral shocks and leads to a shallower gap (their Figure 18),
which only increases the need for dust depletion within the gaps. Finally, we do
not explicitly simulate the MHD turbulence that is believed to provide the
effective viscous stresses in protoplanetary disks (Balbus \& Hawley 1991).
But it appears from previous simulations that adopting the usual $\alpha$
prescription gives results in broad agreement with MHD simulations when
considering gap formation and giant planet migration (Nelson \& Papaloizou 2003;
Winters, Balbus \& Hawley 2003; Papaloizou, Nelson \& Snellgrove 2004). As such,
we do not expect that the qualitative nature of our results have been
compromised by the neglect of the above physical processes.

\section{Summary}
In this paper we present 2-D hydrodynamic simulations to explore
the possibility that the properties of pre/transitional disks are due
to gap opening by planets in viscous disks.  With an improved inner boundary condition,
we found that the surface density of the disk interior to the planet-formed
gap is depleted by the same factor as the mass accretion rate through the
gap and into the inner disk is reduced.  Thus, the substantial accretion
rates of pre/transitional disks require substantial gas surface densities
inside of planet-driven gaps.
We also found that even multiple planets have difficulty in making large
disk gaps and holes which are deep enough to explain the observations.  Thus small
grain depletion seems to be an essential part of the explanation of the
structure of pre/transitional disks, and this may be explained by the dust growth or
the accretion of grain-depleted gas from large disk radii at late
times and dust filtration.  In addition, multiple planets might
account for the inferred structure within the gaps of pre/transitional
disks, in which the inner gaps are much less cleared of small dust than
outer gap regions.

\acknowledgments
This project was initiated during the research programme Dynamics of Discs and
Planets hosted by the Isaac Newton Institute in 2009.
This work was supported in part by NASA grant NNX08AI39G from the Origins
of Solar Systems program, and in part by the University of Michigan.
C.~E. was supported by the National Science Foundation under Award No. 0901947.
Thanks again to Jeremy Hallum for maintaining the compute cluster on which
these simulations were performed.

\clearpage

\begin{deluxetable}{l c c c c}
\tabletypesize{\scriptsize}
\tablewidth{0pt}
\tablecaption{Stellar and Model Properties\\ of LkCa~15, UX~Tau~A,  Rox~44, \& GM Aur from Espaillat et al. 2010 \label{tab:prop}}
\startdata
\hline
\hline
\colhead{} & \colhead{LkCa~15}  & \colhead{UX~Tau~A} & \colhead{Rox~44} & \colhead{GM Aur} \\
\hline
\hline
\multicolumn{5}{c}{Stellar Properties}\\
M$_{*}$ (M$_{\sun}$) & 1.3  & 1.5 & 1.3 & 1.1\\
T$_{*}$ (K) & 4730  & 5520 & 4730 & 4350 \\
$\mdot$ (M$_{\sun}$ yr$^{-1}$) & 3.3$\times$10$^{-9}$ & 1.1$\times$10$^{-8}$ & 9.3$\times$10$^{-9}$ & 7.2$\times$10$^{-9}$ \\
\hline
\multicolumn{5}{c}{Optically Thick Inner Wall}\\
R$_{wall}^i$ (AU) & 0.15 & 0.15  & 0.25 & ...\\
\hline
\multicolumn{5}{c}{Optically Thick Inner Disk}\\
R$_{inner~disk}$ (AU) & $<$0.19  & $<$0.21 & $<$0.4 & ...\\
M$_{inner~disk}$ (M$_{\sun}$) & $<$2$\times$10$^{-4}$ & $<$6$\times$10$^{-5}$ & $<$8$\times$10$^{-5}$ & ...\\
\hline
\multicolumn{5}{c}{Optically Thick Outer Wall} \\
R$_{wall}^o$ (AU) & 58 & 71 & 36 & 20\\
\hline
\multicolumn{5}{c}{Optically Thick Outer Disk}\\
$\epsilon$ & 0.001  & 0.001 & 0.01 & 0.5\\
$\alpha$ & 0.0005  & 0.004 & 0.006 & 0.002\\
M$_{disk}$ (M$_{\sun}$) & 0.1 & 0.04 &0.03 &0.16\\
\hline
\multicolumn{5}{c}{Optically Thin Dust in Gap$/$Hole}\\
M (M$_{\sun}$) & 2$\times$10$^{-11}$  & ... & 2$\times$10$^{-11}$ & 2$\times$10$^{-12}$\\
a$_{min}$ ({\micron}) & 0.005  & ... & 0.005 & 0.005\\
a$_{max}$ ({\micron}) & 0.25  & ... & 0.25 & 0.25\\
R$_{out, thin}$ (AU) & 4 & ...  & 2 & 1
\enddata
\end{deluxetable}

\clearpage

\begin{table}
\begin{center}
\caption{Models \label{tab1}}
\begin{tabular}{ccccccc}

\tableline\tableline
case name    & R$_{in}$ & R$_{out}$ & $\Sigma_{1 AU}$ & R$_{P}$\tablenotemark{a} & M$_{P}$\tablenotemark{b} & Acc \tablenotemark{c} \\
           & AU       & AU        & g cm$^{-2}$       & AU      & M$_{J}$ & f \\
\tableline
PN & 1.5 & 200 & 178 & -  & - & -    \\
P1AN & 1.5 & 200 & 178 & 20  & 1 & No    \\
P1A100 & 1.5 & 200 & 178 & 20  & 0.1 & 100   \\
P1A10 & 1.5 & 200 & 178 & 20  & 0.1 & 10    \\
P1A1 & 1.5 & 200 & 178 & 20  & 0.1 & 1    \\
P2AN & 0.75 & 200 & 178 & 12.5/20  & 1 & No  \\
P2A100 & 0.75 & 200 & 178 & 12.5/20  & 0.1 & 100   \\
P2A10 & 0.75 & 200 & 178 & 12.5/20  & 0.1 & 10    \\
P2A1 & 0.75 & 200 & 178 & 12.5/20  & 0.1 & 1    \\
P3AN & 0.5 & 200 & 178 & 7.5/12.5/20  & 1 & No   \\
P3A100 & 0.5 & 200 & 178 & 7.5/12.5/20  &0.1 & 100   \\
P3A10 & 0.5 & 200 & 178 & 7.5/12.5/20  & 0.1 & 10    \\
P3A1 & 0.5 & 200 & 178 & 7.5/12.5/20  & 0.1 & 1  \\
P4AN & 0.5 & 200 & 178 & 5/7.5/12.5/20  & 1 & No \\
P4A100 & 0.5 & 200 & 178 & 5/7.5/12.5/20  & 0.1 & 100  \\
P4A10 & 0.5 & 200 & 178 & 5/7.5/12.5/20  & 0.1 & 10   \\
P4A1 & 0.5 & 200 & 178 & 5/7.5/12.5/20  & 0.1 & 1   \\
 \tableline
\end{tabular}
\tablenotetext{a}{The initial planet radii. The radii can change with time due to migration and planet-planet interaction.}
\tablenotetext{b}{The initial planet mass. If Acc is not 'No', the mass can increase with time due to accretion from the disk.}
\tablenotetext{c}{The accretion coefficient f is defined in Eqs 4 \& 5.}
\end{center}
\end{table}
\clearpage

\end{document}